\documentclass[aps,pra,twocolumn,superscriptaddress]{revtex4}
\usepackage{amsfonts,amssymb,graphicx}
\usepackage[usenames,dvipsnames]{xcolor}
\usepackage{pdfpages}
\usepackage{epstopdf}
\usepackage{footmisc}
\usepackage[normalem]{ulem}
\usepackage{color}

\begin{document}
\title{Exact zero modes in a quantum compass chain under inhomogeneous transverse fields}
\author{Ning Wu}
\affiliation{Center for Quantum Technology Research, School of Physics, Beijing Institute of Technology, Beijing 100081, China}
\author{Wen-Long You}
\email{youwenlong@gmail.com}
\affiliation{College of Science, Nanjing University of Aeronautics and Astronautics, Nanjing, 211106, China}
\affiliation{School of Physical Science and Technology, Soochow University, Suzhou, Jiangsu 215006, China}

\begin{abstract}
We study the emergence of \emph{exact Majorana zero modes} (EMZMs) in a one-dimensional quantum transverse compass model with both the nearest-neighbor interactions and transverse fields varying over space. By transforming the spin system into a quadratic Majorana-fermion model, we derive an exact formula for the number of the emergent EMZMs, which is found to depend on the \emph{partition nature} of the lattice sites on which the magnetic fields vanish. We also derive explicit expressions for the wavefunctions of these EMZMs and show that they indeed depend on fine features of the foregoing partition of site indices. Based on the above rigorous results about the EMZMs, we provide an interpretation for the interesting dependence of the eigenstate-degeneracy on the transverse fields observed in prior literatures. As a special case, we employ a plane-wave ansatz to exactly solve an open compass chain with alternating nearest-neighbor interactions and staggered magnetic fields. Explicit forms of the canonical Majorana modes diagonalizing the model are given even for finite chains. We show that besides the possibly existing EMZMs, no \emph{almost} Majorana zero modes exist unless the fields on both the two sublattices are turned off. Our results might shed light on the control of ground-state degeneracies by solely tuning the external fields in related systems.
\end{abstract}

\maketitle
\section{Introduction}
\par Recently, there have been extensive studies on the realization of Majorana zero modes in different physical settings (see Refs.~\cite{RPP2012,RMP2015,NPJ2015,Ramon2017} and references therein) because of their potential applications in topological quantum computation. Among these, the emergence of Majorana zero modes in low-dimensional fermionic or spin \emph{lattice} systems has attracted much attention~\cite{Kitaev2001,Fisher2011,Shen2011,Loss2012,Niu2012,PLA2012,Fendley,Sarma2013,Zvyagin2013,Kao2014,Nagaosa2014,Greiter2014,Huang2015,Affleck2015,Fendley2016,Apollaro2016,natphy2016,Mila2016,Mila2017,PRB2017,Schuricht2019}. In a seminal work, Kitaev showed that in a one-dimensional lattice fermion model with nearest-neighbor $p$-wave pairing and open boundaries, a pair of Majorana zero modes can occur at the two ends of the chain when the system is in its topological phase~\cite{Kitaev2001}. In general, these two modes are ``almost" Majorana zero modes (AMZM) in the sense that they decay exponentially away from the edges and possesses an exponentially small excitation energy in the thermodynamic limit. However, exact Majorana zero modes (EMZMs) that strictly commute with the Hamiltonian can emerge~\cite{Fendley, PRB2017} in the special case with equal hopping and pairing strength. These two unpaired EMZMs are spatially separated and lead to two-fold degenerate ground states robust under fermion-parity-preserving perturbations.
\par The appearance of EMZMs is important since they imply the existence of a degenerate ground-state manifold in which quantum information can be stored~\cite{NPJ2015}. The manipulation of EMZMs is thus essential to the realization of a topological quantum computer~\cite{Pachos2017}. However, it is unlikely to directly observe Majorana zero modes in ordinary metals, and nonstandard systems with special properties are necessary for the emergence of EMZMs as nontrivial excitations. It is proposed in Ref.~\cite{Zvyagin2013} that direct observation of edge Majorana fermions is possible in quantum chains. Among these, low-dimensional spin systems have the advantage that some of them can be simulated using nuclear magnetic resonance~\cite{NMR1,NMR2}, trapped ions~\cite{ION}, superconducting quantum circuits~\cite{Youjq,Nori2014}, and optical systems~\cite{Optical}.
\par In this work, we focus on the one-dimensional quantum compass model under inhomogeneous transverse fields as a platform to realize Majorana zero modes. We choose one-dimensional systems here because of their exact solvability and high relevance to the experimental realization of various quasi-one-dimensional quantum materials in recent years. The quantum compass model was initially coined as a minimal model for the exchange interactions in Mott insulators~\cite{Fradkin2005,Dorier2005} and now appears in cross fields~\cite{Feng07,Steinigeweg16,Jafari17,Agrapidis18,Vimal18,Wang18,You2017,You2018,Zotos2018,Katsura2019}. The compass systems are notorious for real-space directional character of the Ising-type exchange terms, and the local $Z_2$ symmetries give rise to many anomalous physical phenomena, such as macroscopic degeneracy~\cite{Brzezicki07,You08,Brzezicki13,You10}. For example, it has been known for a long time that the ground state of the periodic quantum compass chain in the absence of magnetic fields is $2^{L/2-1}$-fold degenerate~\cite{You08}, where $L$ is the number of spins in the chain. However, the huge dimensionality of the ground-state manifold is fragile and gets destroyed by an infinitesimal uniform transverse field~\cite{Sun09,You14}. It was realized very recently that the eigenstate-degeneracy can be restored if the inhomogeneous fields at a portion of the lattice sites vanish~\cite{You2018}. It might be surprising that each energy level is still $2^{L/2-1}$-fold degenerate if the fields on one of the sublattices are switched on~\cite{You2018}. The details of the behavior of the eigenstate-degeneracy, however, will depend on specific configurations of the vanishing fields (see Sec.~\ref{examplesDG} below). In principle, the dependence of the eigenstate-degeneracy on the inhomogeneous fields can be revealed by constructing symmetry operators that commute with the Hamiltonian~\cite{Brzezicki13,You10,You14,You2018}. Although this procedure works for some limiting cases in which simple symmetry operators can be found~\cite{You14}, there is in general no systematic method to construct these operators in a neat way.
\par Motivated by the above observations, we explore the emergence of EMZMs in the quantum compass chain with both the nearest-neighbor interactions and the transverse fields varying over space. By expressing the spin system in terms of Majorana fermions via the Jordan-Wigner transformation, we derive a formula for the number of EMZMs, $\mathcal{N}$, when the magnetic fields on a subset $\mathcal{S}$ of the lattice sites are turned off. It turns out that $\mathcal{N}$ precisely depends on how $\mathcal{S}$ is split into a bunch of \emph{consecutive sequences}. Furthermore, the wavefunctions of these emergent EMZMs are found to be nonlocal for \emph{successive} consecutive sequences with \emph{odd} lengths. These rigorous results on the EMZMs offers a transparent interpretation to the eigenstate-degeneracy problem mentioned above, and the obtained wavefunctions for the EMZMs can be used to construct symmetry spin operators via the inverse Jordan-Wigner transformation of the Majorana fermions. As a consequence, the dependence of the eigenstate-degeneracy on the inhomogeneous field in the transverse compass chain is completely revealed.
\par We also investigate the emergence of almost Majoranan zero modes in an open compass chain with alternating nearest-neighbor interactions and staggered transverse fields. Such a special model is quasi-periodic in the bulk under shifts by four sites in the Majorana representation. By employing a plane-wave ansatz, we are able to analytically diagonalize the model and derive explicit expressions for the wavefunctions of the canonical Majorana fermions in finite systems. It is shown that besides the possibly existing EMZMs, two almost Majorana edge modes also appear provided the fields on both the odd and the even sublattices are turned off.
\par The rest of the paper is organized as follows. In Sec.~\ref{SecII}, we introduce our model and describe its diagonalization in the Majorana representation. In Sec.~\ref{SecIII}, we will investigate the emergence of exact Majorana zero modes in the system in detail. We first outline the main results of this work and summarize them in a Theorem, and then prove the Theorem with the help of four Propositions. In Sec.~\ref{SecIV}, we will briefly review the properties of eigenstate-degeneracy in the inhomogeneous compass chain observed in previous studies, and then give a complete interpretation in terms of the EMZMs. In Sec.~\ref{SecIV} we give the exact solution of an open compass chain with alternating transverse fields. Conclusions are drawn in Sec.~\ref{SecVI}.
\section{Model and diagonalization}\label{SecII}
\subsection{The inhomogeneous compass model and its Majorana representation}
\par The Hamiltonian of a one-dimensional quantum compass model with both nearest-neighbor interactions and transverse fields varying over space can be written as 
\begin{eqnarray}\label{Hobcih}
H_{\mathrm{OBC}}&=&-\sum^{\frac{L}{2}}_{j=1}J^x_j\sigma^x_{2j-1}\sigma^x_{2j}-\sum^{\frac{L}{2}-1}_{j=1}J^y_j\sigma^y_{2j}\sigma^y_{2j+1} \nonumber\\
&& -\sum^L_{j=1}h_j \sigma^z_j,
\end{eqnarray}
for open boundary conditions (OBCs), and as
\begin{eqnarray}\label{Hpbcih}
H_{\mathrm{PBC}}=H_{\mathrm{OBC}}- J^y_{\frac{L}{2}}\sigma^y_{L}\sigma^y_{1}
\end{eqnarray}
for periodic boundary conditions (PBCs). Here, $\sigma^\alpha_l$ are the usual Pauli operators on site $l$, $J^x_j$ ($J^y_j$) is the interaction strength between the two nearest-neighboring sites $2j-1$ and $2j$ ($2j$ and $2j+1$), and $h_l$ is the magnetic field experienced by the spin on site $l$. We choose the number of lattice sites $L$ even since the PBC is not well defined for odd $L$. We assume that $J^x_j$ and $J^y_j$ are all \emph{nonzero}, while a subset of the inhomogeneous fields on certain sites can vanish. We note that inhomogeneous effects can play important roles in the emergence of Majorana zero modes in spin-ladder systems~\cite{Loss2012,Schuricht2019}.
\par Below we mainly focus on the case of OBC and will address the case of PBC when discussing the eigenstate-degeneracy of the model. The inhomogeneous compass chain described by Eq.~(\ref{Hobcih}) can be viewed as a generalization of the celebrated Kitaev honeycomb model~\cite{Feng07,Kitaev2006} in the one-dimensional limit by introducing spatially varying nearest-neighboring interactions and transverse fields. Using the Jordan-Wigner transformation
\begin{eqnarray}\label{JW}
 \sigma^x_j-i\sigma^y_j  =2a_j e^{i\pi\sum^{j-1}_{l=1}a^\dag_la_l},~\sigma^z_j=2a^\dag_ja_j-1,
\end{eqnarray}
where $a^\dag_j$ is a fermonic creation operator on site $j$, the Hamiltonian $H_{\mathrm{OBC}}$ can be mapped to a spinless fermion model described by
\begin{figure}
\includegraphics[width=.52\textwidth]{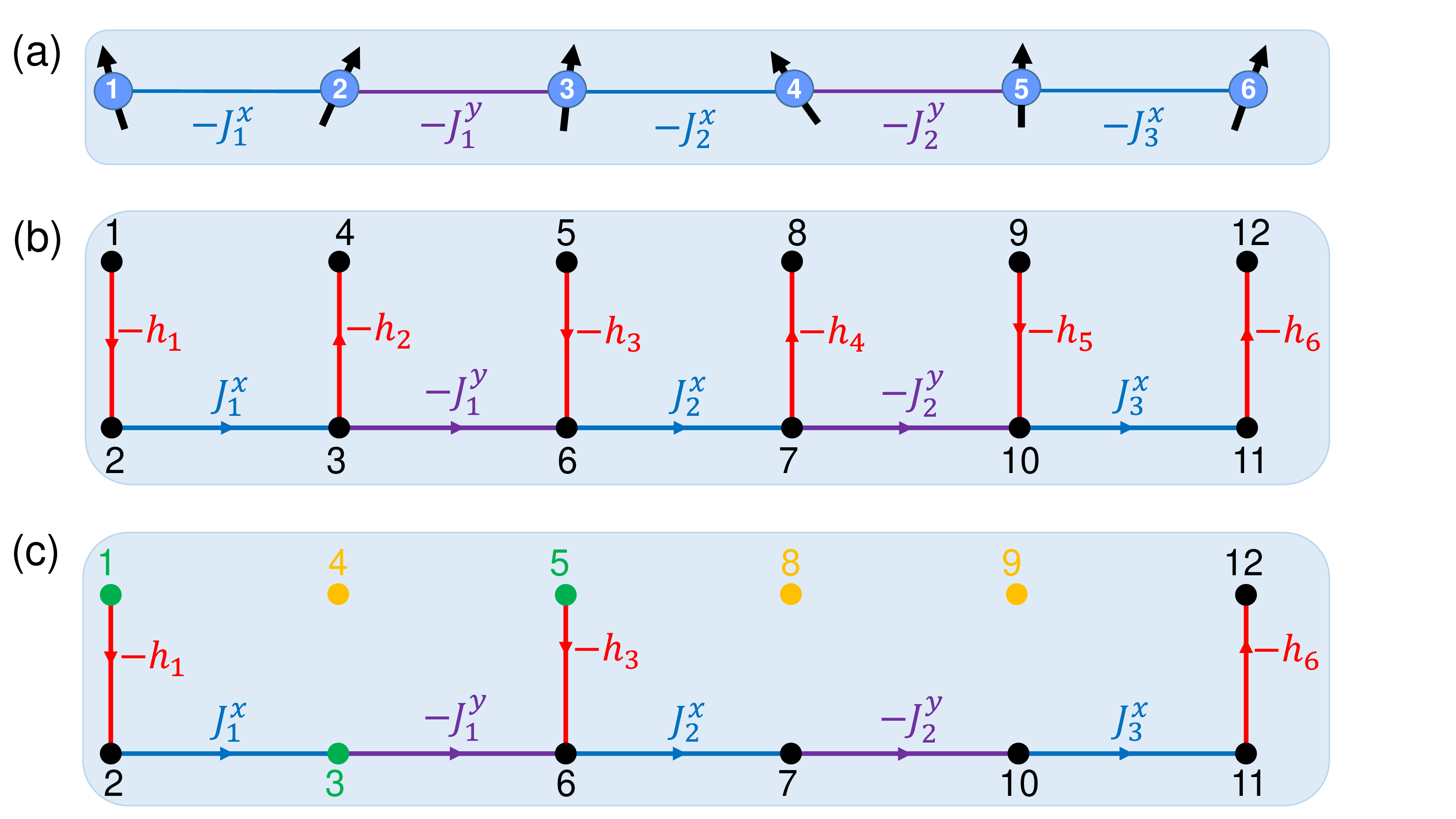}
\caption{(a) Real-space representation of an open transverse compass chain with $L=6$ spins. (b) The corresponding snake chain representation for $H^{(\mathrm{MF})}_{\mathrm{OBC}}$ in the Majorana space. Finite interactions between different MFs form a comb-like structure with $L$ teeth. (c) Turning off the fields on sites $\mathcal{S}=\{2,4,5\}$ gives rise to three isolated EMZMs $d_4$, $d_8$, and $d_9$ (orange points), as well as a nontrivial EMZM that is a linear superposition of $d_1$, $d_3$, and $d_5$ (green points).}
\label{Fig1}
\end{figure}
\begin{eqnarray}\label{HJW}
H^{(\mathrm{JW})}_{\mathrm{OBC}}&=&-\sum^{\frac{L}{2}}_{j=1}J^x_j[(a^\dag_{2j-1} a_{2j}-a_{2j-1}a_{2j})+\mathrm{H.c.}]\nonumber\\
&&-\sum^{\frac{L}{2}-1}_{j=1}J^y_j[(a^\dag_{2j} a_{2j+1}+a_{2j}a_{2j+1})+\mathrm{H.c.}]\nonumber\\
&&-\sum^L_{j=1}h_j (2a^\dag_ja_j-1),
\end{eqnarray}
where $\mathrm{H.c.}$ denotes Hermitian conjugate.
In contrast to Kitaev's $p$-wave superconductor model that can be mapped from the quantum Ising chain via the Jordan-Wigner transformation~\cite{Fendley}, in the fermionic chain described by $H^{(\mathrm{JW})}_{\mathrm{OBC}}$ the relative sign between the nearest-neighboring hopping term and pairing term on bond $(l,l+1)$ depends on the parity of $l$.
\par We now define $2L$ Majorana fermions (MFs)
\begin{eqnarray}\label{MF}
\left(
  \begin{array}{c}
    d_{2j-1} \\
    d_{2j} \\
  \end{array}
\right)=\Gamma_2\left(
  \begin{array}{c}
    a_j \\
    a^\dag_j \\
  \end{array}
\right),~~\Gamma_2\equiv\left(
                    \begin{array}{cc}
                      1 & 1 \\
                      -i & i \\
                    \end{array}
                  \right),
\end{eqnarray}
for $j=1,2,\cdots,L$. Note that $\Gamma^\dag_2\Gamma_2=\Gamma_2\Gamma^\dag_2=2$, and the MFs satisfy the canonical anticommutation relations $\{d_l,d_{l'}\}=2\delta_{l,l'}$. In terms of the MFs, $H^{(\mathrm{JW})}_{\mathrm{OBC}}$ can be rewritten in the Majorana representation as
\begin{eqnarray}\label{HMFDD}
H^{(\mathrm{MF})}_{\mathrm{OBC}}&=& \frac{i}{4}\mathcal{D}^T\mathcal{H}_{\mathrm{OBC}}\mathcal{D},\nonumber\\
\mathcal{D}&=&(d_1,d_2,\cdots,d_{2L})^T,
\end{eqnarray}
where $T$ denotes the matrix transpose and $\mathcal{H}_{\mathrm{OBC}}$ is a $2L\times 2L$ real antisymmetric matrix of the form
\begin{widetext}
\begin{eqnarray}
&&\mathcal{H}_{\mathrm{OBC}}=2\left(
                \begin{array}{cccccccccccccc}
                  0 & -h_1 &   &   &   &   &   &  & &      &     & &       \\
                  h_1 & 0 & J^x_1 &   &   &   &   &  &    &      & &   &      \\
                   & -J^x_1 & 0 & -h_2 & 0 & -J^y_1 &     &     & &  &   &   &      \\
                   &   & h_2 & 0  & 0  & 0  &   &  &   &     &    &  &     \\
                   &   &  0 & 0  &  0 & -h_3 &   &  &   &   &      &  &      \\
                    &   & J^y_1 & 0 & h_3 & 0 & J^x_2 &   &     & & &   &    \\
                    &   &   &   &   & -J^x_2 & 0 & -h_4   &   &    &    &   &     \\
                    &  &   &  &   &   &h_4   & \ddots  & &    &    &   &     \\
                     &  &   &  &   &   &   &    &\ddots &    &  \vdots  &   &     \\
                            &  &   &  &   & &  &    &  &   \ddots  & -h_{L-1}     &   &     \\
                &  &  &  &     &   &  &  &    \ldots   & h_{L-1} & 0 & J^x_{\frac{L}{2}} &    \\
                    &  & &      &   &  &  &    &  &   & -J^x_{\frac{L}{2}} & 0 & -h_L   \\
                    &  & &      &   &  &   &     &   &   &   & h_L & 0   \\
                \end{array}
              \right).
\end{eqnarray}
\end{widetext}
\par We arrange the $2L$ MFs into a ``Majorana snake chain"~\cite{PLA2012} as shown in Fig~\ref{Fig1}(a), where finite interactions between different MFs in Eq.~(\ref{HMFDD}) form an \emph{$L$-tooth comb} with the field $h_l$ residing on the $l$th tooth.
As a traceless antisymmetric matrix of even dimension, $\mathcal{H}_{\mathrm{OBC}}$ can be brought into the following form
\begin{eqnarray}\label{Lamb}
&&\tilde{\mathcal{H}}_{\mathrm{OBC}}=\left(
            \begin{array}{ccccccc}
              0 & \epsilon_1 &   &   &   &   &   \\
              -\epsilon_1  & 0 &   &  &   &   &   \\
                &  & \cdot &  &   &   &   \\
                &   &   & \cdot &   &   &   \\
                &   &   &   & \cdot &   &   \\
                &   &   &   &   & 0 & \epsilon_L \\
                &   &   &   &   & -\epsilon_L & 0 \\
            \end{array}
          \right)
\end{eqnarray}
through the orthogonal transformation
\begin{eqnarray}\label{WWT}
\tilde{\mathcal{H}}_{\mathrm{OBC}}=W\mathcal{H}_{\mathrm{OBC}}W^T,
\end{eqnarray}
where $W$ is a $2L\times 2L$ real orthogonal matrix and $\epsilon_i\geq 0$. It is easy to see that the eigenvalues of $\tilde{\mathcal{H}}_{\mathrm{OBC}}$, and hence of $\mathcal{H}_{\mathrm{OBC}}$, always appear in pairs and are given by $\{\pm i\epsilon_i\}$. Combining Eq.~(\ref{WWT}) with Eq.~(\ref{HMFDD}), $H^{(\mathrm{MF})}_{\mathrm{OBC}}$ can be reduced to the canonical form~\cite{Kitaev2001}
\begin{eqnarray}\label{canon}
H^{(\mathrm{MF})}_{\mathrm{OBC}}= \frac{i}{4}\mathcal{B}^T\tilde{\mathcal{H}}_{\mathrm{OBC}}\mathcal{B}=\frac{i}{4}\sum^L_{j=1}\epsilon_j(b'_jb''_j-b''_j b'_j),
\end{eqnarray}
where
\begin{eqnarray}\label{bb}
\mathcal{B}\equiv(b'_1,b''_1,\cdots,b'_L,b''_L)^T=W\mathcal{D}
\end{eqnarray}
is another set of MFs due to the orthogonality of $W$.
\par \par If we further define the ordinary fermions
\begin{eqnarray}\label{ff}
 \left(
   \begin{array}{c}
     f_j \\
     f^\dag_j \\
   \end{array}
 \right)
  =\frac{1}{2}\Gamma^\dag_2 \left(
   \begin{array}{c}
     b'_j \\
     b''_j \\
   \end{array}
 \right),
\end{eqnarray}
then ${H}^{\mathrm{(MF)}}_{\mathrm{OBC}}$ can finally be diagonalized as
\begin{eqnarray}\label{HOBCff}
H^{(\mathrm{F})}_{\mathrm{OBC}}=\sum^L_{j=1}\epsilon_j\left(f^\dag_j f_j-\frac{1}{2}\right).
\end{eqnarray}
We see that finding out the real orthogonal matrix $W$ that brings $\mathcal{H}_{\mathrm{OBC}}$ into $\tilde{\mathcal{H}}_{\mathrm{OBC}}$ is equivalent to the diagonalization of the spin model $H_{\rm{OBC}}$. Physically, the $(2j-1)$th [the $(2j)$-th] row of $W$, $W_{2j-1,:}$ ($W_{2j,:}$),
 characterizes the spreading of the MF $b'_j$ ($b''_j$) in the real-space representation of the snake chain, and will be referred to as the \emph{wavefunction} of $b'_j$ ($b''_j$).
\par Below we are interested in the eigen-problem
\begin{eqnarray}\label{eigen}
\mathcal{H}_{\mathrm{OBC}}V=2\lambda V,
\end{eqnarray}
where the factor 2 is introduced for later convenience. The quasi-periodic structure of $\mathcal{H}_{\mathrm{OBC}}$ in the bulk suggests us to write the eigenvector $V$ as
\begin{eqnarray}\label{VABCD}
V=\left(A_1,B_1,C_1,D_1,\cdots,A_{\frac{L}{2}},B_{\frac{L}{2}},C_{\frac{L}{2}},D_{\frac{L}{2}}\right)^T.
\end{eqnarray}
Using the components $A_i$, $B_i$, $C_i$, and $D_i$, Eq.~(\ref{eigen}) can be explicitly written out as $2L-2$ \emph{bulk equations}:
\begin{eqnarray}
\label{bulk0_1}
-h_{2j-1} B_j&=&\lambda A_j,~~ h_{2j}C_j=\lambda D_j,~\forall j
\end{eqnarray}
and
\begin{eqnarray}
\label{bulk0_3}
-J^x_jB_j-h_{2j}D_j-J^y_jB_{j+1}&=&\lambda C_j,\\
\label{bulk0_4}
J^y_jC_j+h_{2j+1}A_{j+1}+J^x_{j+1}C_{j+1}&=&\lambda B_{j+1},
\end{eqnarray}
for $j=1,2,\cdots,\frac{L}{2}-1$, as well as two \emph{boundary equations}
\begin{eqnarray}
\label{boud0_1}
-J^x_{\frac{L}{2}}B_{\frac{L}{2}}-h_{L}D_{\frac{L}{2}}&=&\lambda C_{\frac{L}{2}},\\
\label{boud0_2}
h_{1}A_{1}+J^x_1C_{1}&=&\lambda B_{1}.
\end{eqnarray}
\subsection{Relation between the eigenvectors of $\mathcal{H}_{\mathrm{OBC}}$ and the matrix $W$}\label{relation}
\par Before ending this section, let us discuss the relationship between the eigenvector $V$'s of $\mathcal{H}_{\mathrm{OBC}}$ and the matrix $W$. In general, the $V$'s are not consistent with the wavefunctions for the MFs $\{b'_j,b''_j\}$ since the matrix $W$ only brings $\mathcal{H}_{\mathrm{OBC}}$ into the canonical form $\tilde{\mathcal{H}}_{\mathrm{OBC}}$. Remember that the $2L$ eigenvalues of $\mathcal{H}_{\mathrm{OBC}}$ always appear in pairs $\{\pm 2\lambda_j\}$, so we can arrange them in the order
\begin{eqnarray}\label{Theta}
\Theta&\equiv&\mathrm{diag}(2\lambda_1,-2\lambda_1,\cdots,2\lambda_L,-2\lambda_L).
\end{eqnarray}
According to the spectral theorem, the corresponding eigenvectors $\{V_{+,j},V_{-,j}\}$ can be appropriately chosen to form a unitary matrix
\begin{eqnarray}
X\equiv (V_{+,1},V_{-,1},\cdots,V_{+,L},V_{-,L}),
\end{eqnarray}
so that
\begin{eqnarray}\label{HX}
\mathcal{H}_{\mathrm{OBC}}X=X\Theta.
\end{eqnarray}
We define the $2L\times 2L$ matrix
\begin{eqnarray}
\Gamma_{2L}\equiv\oplus^L_{j=1} \Gamma_{2}.
\end{eqnarray}
It is easy to check that
\begin{eqnarray}
\Gamma_{2L}\Gamma^\dag_{2L}=\Gamma^\dag_{2L}\Gamma_{2L}=2.
\end{eqnarray}
From Eq.~(\ref{MF}) and Eq.~(\ref{ff}), we get
\begin{eqnarray}
(f_1,f^\dag_1,\cdots,f_{L},f^\dag_L)^T=\frac{1}{2}\Gamma^\dag_{2L}W\mathcal{D}.
\end{eqnarray}
To find the relation between $X$ and $W$, we use the above equation to rewrite $\mathcal{H}^{\mathrm{(F)}}_{\mathrm{OBC}}$ given by Eq.~(\ref{HOBCff}) as
\begin{eqnarray}\label{WD}
H^{\mathrm{(F)}}_{\mathrm{OBC}}
&=&\frac{1}{8}(\Gamma^\dag_{2L}W\mathcal{D})^\dag\Lambda(\Gamma^\dag_{2L}W\mathcal{D})\nonumber\\
&=&\frac{1}{8}\mathcal{D}^T(W^T\Gamma_{2L})\Lambda(\Gamma^\dag_{2L}W)\mathcal{D},
\end{eqnarray}
where $\Lambda=\mathrm{diag} (\epsilon_1,-\epsilon_1,\cdots,\epsilon_L,-\epsilon_L)$ and we have used $\mathcal{D}^T=\mathcal{D}^\dag$ and $W^T=W^\dag$. Compare Eq.~(\ref{WD}) with Eq.~(\ref{HMFDD}), we obtain
\begin{eqnarray}
i\mathcal{H}_{\mathrm{OBC}}=\frac{1}{2}W^T\Gamma_{2L}\Lambda \Gamma^\dag_{2L}W.
\end{eqnarray}
Multiplying both sides of the last equation by $W^T\Gamma_{2L}$, we get
\begin{eqnarray}\label{HWG}
\mathcal{H}_{\mathrm{OBC}}(W^T\Gamma_{2L}) =(W^T\Gamma_{2L})(-i\Lambda).
\end{eqnarray}
By comparing Eq.~(\ref{HWG}) with Eq.~(\ref{HX}) we finally obtain the relation between the two matrices $X$ and $W$
\begin{eqnarray}
\label{Xw}
X&=&\frac{1}{\sqrt{2}}W^T\Gamma_{2L},\\
\Theta&=&-i\Lambda.
\end{eqnarray}
Suppose we already obtain $X$ by solving Eq.~(\ref{eigen}), then the $W$ matrix can be calculated from Eq.~(\ref{Xw}) as
\begin{eqnarray}\label{getW}
W&=&\frac{1}{\sqrt{2}}\Gamma^*_{2L}X^T
                                         =\frac{1}{\sqrt{2}}\left(
                                                           \begin{array}{c}
                                                             V^T_{+,1}+V^T_{-,1} \\
                                                             i\left(V^T_{+,1}-V^T_{-,1}\right) \\
                                                             \vdots \\
                                                             V^T_{+,L}+V^T_{-,L} \\
                                                             i\left(V^T_{+,L}-V^T_{-,L}\right) \\
                                                           \end{array}
                                                         \right).
\end{eqnarray}
The reality of the matrix $W$ enforces us to choose
\begin{eqnarray}\label{VVstar}
V_{+,j}=V^{*}_{-,j},~\forall j.
\end{eqnarray}
Therefore, the wavefunctions for the two MFs $b'_j$ and $b''_j$ can be written as
\begin{eqnarray}
W_{2j-1,:}&=&\frac{1}{\sqrt{2}}(V^T_{+,1}+V^T_{-,1}),\nonumber\\
W_{2j,:}&=&\frac{i}{\sqrt{2}}(V^T_{+,1}-V^T_{-,1}).
\end{eqnarray}
\par If one of the eigenvalues of $\mathcal{H}_{\mathrm{OBC}}$, say $2\lambda_l$ is zero, then the two eigenvectors $V_{+,l}$ and $V_{-,l}$ become degenerate. In turn, both $W_{2l-1,:}^T$ and $W_{2l,:}^T$ are the eigenvectors of $\mathcal{H}_{\mathrm{OBC}}$ with eigenvalue zero. Thus, a pair of \emph{real} zero-energy eigenvectors orthogonal to each other are usually consistent with the wavefunctions of the corresponding MFs with a zero energy. These properties of the eigenvectors of $\mathcal{H}_{\mathrm{OBC}}$ will be used in the next section to identify the wavefunctions for the EMZMs. It should be mentioned that the wavefunctions for the zero-energy MFs are not unique.
\section{Emergence of exact Majorana zero modes}\label{SecIII}
\subsection{Main results}
\par As suggested by Fendley~\cite{Fendley}, an EMZM $\eta$ is defined as an operator satisfying the following three properties:
\par i) $\eta$ commutes with $H^{(\mathrm{MF})}_{\mathrm{OBC}}$,
\par ii) $\eta$ anticommutes with the fermion parity operator $P=e^{i\pi\sum^L_{j=1}a^\dag_j a_j}=\prod^L_{j=1}(-id_{2j-1}d_{2j})$,
\par iii) $\eta$ is normalizable.
\par According to the above definition, it is apparent that the two MFs $(b'_j,b''_j)$ are a pair of EMZMs if $\epsilon_j=0$: Firstly, $[H^{(\mathrm{MF})}_{\mathrm{OBC}},b'_j]=[H^{(\mathrm{MF})}_{\mathrm{OBC}},b''_j]=0$ is obviously fulfilled as $b'_j$ and $b''_j$ do not appear in $H^{(\mathrm{MF})}_{\mathrm{OBC}}$; secondly, one can easily check that $\{b'_j,P\}=\{b''_j,P\}=0$ since both $b'_j$ and $b''_j$ are linear combinations of $\{d_l\}$; thirdly, as MFs, both $b'_j$ and $b''_j$ square to 1.
\par We call the MFs $(b'_j,b''_j)$ a pair of \emph{almost Majorana zero modes} if their corresponding single-particle energy $\epsilon_j(L)$ decays exponentially (faster than $\sim 1/L$) as $L\to\infty$. Correspondingly, for the AMZMs the condition i) above should be loosened as $[H^{(\mathrm{MF})}_{\mathrm{OBC}},b'_j]=\hat{o}'$, $[H^{(\mathrm{MF})}_{\mathrm{OBC}},b''_j]=\hat{o}''$, where $\hat{o}'$ and $\hat{o}''$ are operators exponentially small in the thermodynamic limit $L\to\infty$. Typical examples of AMZMs include Majorana edge zero modes presented in Kitaev's $p$-wave superconducting chain~\cite{Kitaev2001,PRB2017} and the quantum Ising chain with open boundaries~\cite{Fendley} when the system belongs to the topological phase. Two spatially separated AMZMs are able to form a Dirac fermion that can serve as a topological qubit protected against the coupling between a single zero-mode operator and nonzero-mode operators~\cite{NPJ2015}.
\par  To explore the conditions under which EMZMs can emerge for the Majorana Hamiltonian $H^{(\mathrm{MF})}_{\mathrm{OBC}}$, we set $\lambda=0$ in Eqs.~(\ref{bulk0_1})-(\ref{boud0_2}) to get
\begin{eqnarray}
\label{nbulk0_1}
-h_{2j-1} B_j=0,~~ h_{2j}C_j&=&0,~\forall j \\
\label{nbulk0_3}
-J^x_jB_j-h_{2j}D_j-J^y_jB_{j+1}&=&0,~j\neq\frac{L}{2}\\
\label{nbulk0_4}
J^y_jC_j+h_{2j+1}A_{j+1}+J^x_{j+1}C_{j+1}&=&0,~j\neq\frac{L}{2}\\
\label{nboud0_1}
-J^x_{\frac{L}{2}}B_{\frac{L}{2}}-h_{L}D_{\frac{L}{2}}&=&0,\\
\label{nboud0_2}
h_{1}A_{1}+J^x_1C_{1}&=&0.
\end{eqnarray}
Based on Eqs.~(\ref{nbulk0_1})-(\ref{nboud0_2}), we first show
\par\emph{Proposition 1}: The Hamiltonian $H^{(\mathrm{MF})}_{\mathrm{OBC}}$ given by Eq.~(\ref{HMFDD}) does not support any EMZM if $h_l\neq 0,~\forall l$.
\par \emph{Proof}: If none of $h_l$ vanishes, then from Eq.~(\ref{nbulk0_1}) we must have $B_j=C_j=0,~\forall j$, so that $D_j=A_j=0,~\forall j$, according to Eqs.~(\ref{nbulk0_3})-(\ref{nboud0_2}). We thus obtain a unique trivial solution $V=0$.
\par Proposition 1 indicates that the EMZMs can possibly occur if and only if the magnetic fields on a subset of the lattice sites, say
\begin{eqnarray}\label{setS}
\mathcal{S}=\{l_1,l_2,\cdots,l_m\},~~ 1\leq l_1<l_2<\cdots<l_m\leq L,\nonumber
\end{eqnarray}
vanish. Actually, it is easy to observe by investigating Fig.~\ref{Fig1}(b) that there are $m$ MFs isolated from the comb after the teeth with indices drawn from $\mathcal{S}$ are broken up. In turn, the isolation of these MFs leads to the following $m$ EMZMs
\begin{eqnarray}\label{simpleEMZMs}
d_{2l_i-\xi(l_i)},~i=1,2,\cdots,m.
\end{eqnarray}
where
\begin{equation}\label{eofunc}
\xi(n) =
\cases{
0, & $n=\mathrm{even}$ \cr
1, & $n=\mathrm{odd}$. \cr
}
\end{equation}
\par We now naturally may ask: Are there any other EMZMs besides these $m$ \emph{simple} EMZMs? We answer this question by
\par \emph{Theorem}: Among the $L$ fields $\{h_1,h_2,\cdots,h_L\}$, suppose $\mathcal{S}=\{l_1,l_2,\cdots,l_m\}$ is a set of site indices for which $h_i=0$ if $i\in \mathcal{S}$ and $h_i\neq0$ if $i\notin \mathcal{S}$. The set $\mathcal{S}$ can always be written as the union of $n$ ordered \emph{consecutive sequences} $T_j=(l_{u_j},l_{u_j}+1,\cdots,l_{u_j}+v_j-1)$ with length $v_j$, i.e., $\mathcal{S}=\{T_1,T_2,\cdots,T_n\}$. Then the total number of EMZMs for $H^{(\mathrm{MF})}_{\mathrm{OBC}}$ is
\begin{eqnarray}\label{No}
\mathcal{N}&=&m+\sum^n_{l=1}\xi(v_l).
\end{eqnarray}
\par The above Theorem is the main result of this work and will be proved at a later stage, it shows that the number of EMZMs for $H^{(\mathrm{MF})}_{\mathrm{OBC}}$ is completely determined by the partition nature of the set $\mathcal{S}$. It is easy to see that $m\geq n$ and $u_1=1$, $l_{u_n}+v_n-1=l_m$. Note that $l_{u_{j+1}}\geq l_{u_j}+v_j+1$, otherwise if $l_{u_{j+1}}=l_{u_j}+v_j$, then $T_j$ and $T_{j+1}$ form a single consecutive sequence of length $v_j+v_{j+1}$. We show in Fig.~\ref{Fig1}(b) the example of $L=6$ and $\mathcal{S}=\{2,4,5\}$ with $T_1=(2)$ and $T_2=(4,5)$, giving $\mathcal{N}=3+\xi(1)+\xi(2)=4$ EMZMs according to Eq.~(\ref{No}). This indicates that there exists one more EMZM besides the isolated MFs $d_4,~d_8$, and $d_9$ given by Eq.~(\ref{simpleEMZMs}). We will see later that this extra EMZM is indeed a linear superposition of $d_1$, $d_3$, and $d_5$ [green circles in Fig.~\ref{Fig1}(b)].
\par Before proving the above Theorem, let us look at some of its consequences in several typical cases:
\par 1) If the length of all $T_j$ is 1, then $n=m$ and $v_j=1$, $\forall j$, yielding $2m$ EMZMs. In particular, if all the fields on the odd/even sublattice are turned off, there will be $L$ EMZMs.
\par 2) In the opposite limit with all fields being turned off, we have $m=L$ and $n=1$, but there are still $L$ EMZMs according to Eq.~(\ref{No}).
\subsection{Proof of the Theorem}
\par To prove the Theorem, we start with the simplest case.
\par \emph{Proposition 2}: Suppose $\mathcal{S}$ consists of a single consecutive sequence $T$ of length $v$, i.e., $\mathcal{S}=T=(l_u,l_u+1,\cdots,l_u+v-1)$, then there exist $v$ EMZMs if $v$ is even, and $v+1$ zero modes if $v$ is odd. For even $v$, the $v$ EMZMs are merely the isolated simple ones given by Eq.~(\ref{simpleEMZMs}); for odd $v$, besides the $v$ simple modes, there is an additional EMZM emerging from the diagonalization of partially connected comb.
\par We leave the proof of Proposition 2 to Appendix~\ref{AppA} for the sake of readability. The basic idea consists of repeatedly applying the eigen-equations given by Eqs.~(\ref{nbulk0_1})-(\ref{nboud0_2}) under the condition $h_{l_u}=h_{l_u+1}=\cdots =h_{l_u+v-1}=0$. Let us focus on the case of even $l_u$ as similar results hold for odd $l_u$. For even $v$, there are $v$ simple solutions of the eigenvector $V$ [cf. Eq.~(\ref{evenvV})], which result in $v$ isolated EMZMs $d_{4j}$ and $d_{4j+1}$, $j=\frac{l_u}{2},\frac{l_u}{2}+1,\cdots,\frac{l_u}{2}+\frac{v}{2}-1$. These $v$ EMZMs reside on a segment of length $2v$ along the snake chain.
\par For odd $v$, we have to distinguish the cases with $v=1$ and $v\geq 3$. If $v=1$, then only $h_{l_u}$ is zero, so we have an isolated EMZM $d_{2l_u}$ and a nontrivial EMZM [cf. Eq.~(\ref{Vv=1})] that is a linear superposition of three MFs
\begin{eqnarray}\label{EMZMVv=1}
 \mathcal{J}^{(x,h)}_{\frac{l_u}{2}} d_{2l_u-3}+d_{2l_u-1}+\mathcal{J}^{(y,h)}_{\frac{l_u}{2}}d_{2l_u+1},
\end{eqnarray}
where we have left out a normalization constant and defined
\begin{eqnarray}\label{Jhua}
\mathcal{J}^{(x,h)}_{i}\equiv-\frac{J^x_i}{h_{2i-1}},~~ \mathcal{J}^{(y,h)}_{i}\equiv-\frac{J^y_i}{h_{2i+1}}.
\end{eqnarray}
\par For $v\geq 3$, we get $v$ simple isolated EMZMs $d_{4j}$ $\left(j=\frac{l_u}{2},\frac{l_u}{2}+1,\cdots,\frac{l_u}{2}+\frac{v-1}{2}\right)$ and  $d_{4j+1}$ $\left(j=\frac{l_u}{2},\frac{l_u}{2}+1,\cdots,\frac{l_u+v-3}{2}\right)$, as well as a (unnormalized) nontrivial EMZM [cf. Eq.~(\ref{Vvneq1})]
\begin{eqnarray}
&&\mathcal{J}^{(x,h)}_{\frac{l_u}{2}}d_{2l_u-3}+d_{2l_u-1} +\sum^{\frac{v-3}{2}}_{l=0}\prod^{i+l}_{i=\frac{l_u}{2}}\mathcal{J}^{(y,x)}_{i} d_{2l_u+3+4l}\nonumber\\
&&+\prod^{\frac{l_u+v-3}{2}}_{i=\frac{l_u}{2}}\mathcal{J}^{(y,x)}_{i}\mathcal{J}^{(y,h)}_{\frac{l_u+v-1}{2}}d_{2l_u+2v-1},
\end{eqnarray}
where we further defined $\mathcal{J}^{(y,x)}_i\equiv- J^y_i/J^x_{i+1} $. It is easy to see that the wavefunctions of the foregoing $v$ simple EMZMs and those of the nontrivial EMZM are orthogonal to each other, with the latter spreading over $2v+3$ lattice sites along the snake chain. This indicates that this nontrivial EMZM is highly nonlocal if the single sequence $T$ is long enough.
\par In general, the set $\mathcal{S}$ is a union of multiple consecutive sequences, so we still need to study the situation where more than one consecutive sequence is present.
\begin{figure}
\includegraphics[width=.48\textwidth]{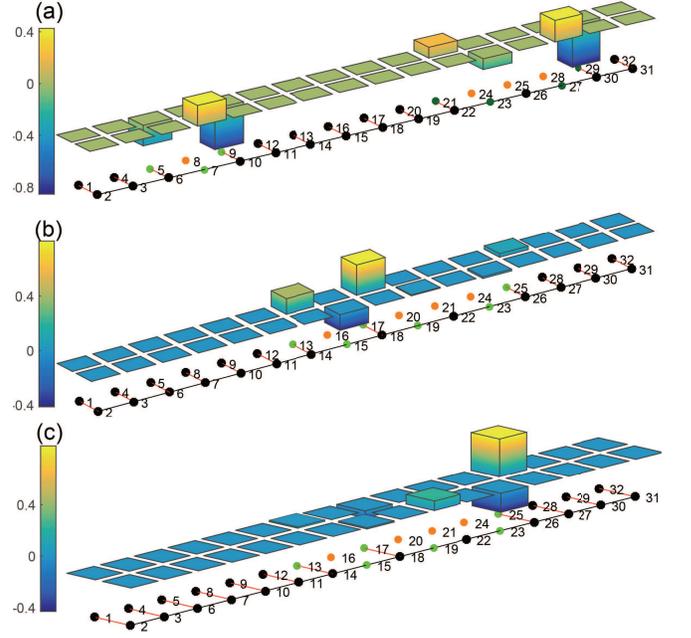}
\caption{(a) Finite components of wavefunctions for the two nontrivial EMZMs with $L=16$ and $\mathcal{S}=\{T_1,T_2\}$. Green: $T_1=(4)$; dark green: $T_2=(12,13,14)$. The orange points denote the four isolated EMZMs given by Eq.~(\ref{simpleEMZMs}). We show the two wavefunctions in the same figure because $T_1$ and $T_2$ are uncoupled. The vertical displacements on top of the snake chain indicate the amplitudes of the individual components (For the sake of brevity, the original spin representation is not shown). (b) and (c): Wavefunctions of the two nontrivial EMZMs generated by two coupled consecutive sequences $T_1=(8)$ and $T_2=(10,11,12)$. Calculations are performed for $J^x_j=1$ and $J^y_j=2$. All the nonvanishing magnetic fields are chosen to be $h_l=1$ ($2$) for odd (even) $l$.
}
\label{Fig2}
\end{figure}
We call two neighboring consecutive sequences $T_i$ and $T_{i+1}$ \emph{successive} if $l_{u_{i+1}}= l_{u_i}+v_i+1$, and \emph{unsuccessive} if $l_{u_{i+1}}\geq l_{u_i}+v_i+2$. As an example, we show in Fig~\ref{Fig2}(a) the wavefunctions of the two EMZMs generated by the two unsuccessive sequences $T_1=(4)$ and $T_2=(12,13,14)$ for $L=16$. It is easy to see that the two wavefunctions are spatially separated and orthogonal to each other.
\par \emph{Proposition 3}: The eigen-equations associated with two neighboring sequences $T_i$ and $T_{i+1}$ are \emph{coupled} only if $T_i$ and $T_{i+1}$ are \emph{successive}, and at the same time \emph{both} $v_i$ and $v_{i+1}$ are odd.
\par \emph{Proof}: By investigating the eigen-equations related to $T=(l_u,l_u+1,\cdots,l_u+v-1)$, we can find out all the possibly \emph{nonvanishing} variables appearing in these equations. As shown in Appendix~\ref{AppA}, for even $v$ we always have $B_j=C_j=0, \forall j$, so there are no such nonzero variables besides those determining the simple EMZMs; For odd $v$, the possibly nonvanishing variables involved are
\begin{eqnarray}
A_{\frac{l_u}{2}}, A_{\frac{l_u+ v+1}{2}},~C_{\frac{l_u}{2}},C_{\frac{l_u}{2}+1},\cdots,C_{\frac{l_u+v-1}{2}},
\end{eqnarray}
for even $l_u$ [cf. equation set (\ref{EOMeoC})], and are
\begin{eqnarray}
D_{\frac{l_u-1}{2}},D_{\frac{l_u+v}{2}},~B_{\frac{l_u+1}{2}},\cdots,B_{\frac{l_u+v}{2}},
\end{eqnarray}
for odd $l_u$ [cf. equation set (\ref{EOMooB})].
\par Hence, the eigen-equations related to $T_i$ and $T_{i+1}$ are never coupled if at least one of $v_i$ and $v_{i+1}$ is even. However, if $T_i$ and $T_{i+1}$ are successive and both $v_i$ and $v_{i+1}$ are odd, then $l_{u_i}$ and $l_{u_{i+1}}$ have the same parity since $l_{u_{i+1}}=l_{u_i}+v_i+1$. For even $l_{u_i}$, the variables related to $T_i$ ($T_{i+1}$) and with the \emph{largest} (\emph{smallest}) indices are  $A_{\frac{l_{u_i}+v_i+1}{2}}$ and $C_{\frac{l_{u_i}+v_i-1}{2}}$ $\left(A_{\frac{l_{u_{i+1}}}{2}}~\mathrm{and}~C_{\frac{l_{u_{i+1}}}{2}}\right)$.  Correspondingly, the eigen-equations for $T_i$ and $T_{i+1}$ will share a common variable $A_{\frac{l_{u_i}+v_i+1}{2}}=A_{\frac{l_{u_{i+1}}}{2}}$. For odd $l_{u_i}$, the variables related to $T_i$ ($T_{i+1}$) and with the largest (smallest) indices are  $D_{\frac{l_{u_i}+v_i}{2}}$ and $B_{\frac{l_{u_i}+v_i}{2}}$ $\left(D_{\frac{l_{u_{i+1}}-1}{2}}~\mathrm{and}~B_{\frac{l_{u_{i+1}}+1}{2}}\right)$.  Correspondingly, the eigen-equations for $T_i$ and $T_{i+1}$ will share a common variable $D_{\frac{l_{u_i}+v_i}{2}}=D_{\frac{l_{u_{i+1}}-1}{2}}$. Hence, Proposition 3.
\par Physically, any sequence with even length $v$ only leads to $v$ simple isolated EMZMs locating on a segment of length $2v$ within the snake chain. However, the nontrivial EMZMs for a single odd sequence spread over a segment with more than $2v$ sites [$2v+3$ for $v\geq 3$ and $3$ for $v=1$, see Eq.~(\ref{Vv=1})], which makes two such successive sequences spatially coupled with each other. We now focus on the case in Proposition 3, and will show
\par \emph{Proposition 4}. Suppose $T_i$, $T_{i+1}$,$\cdots$, and $T_{i+\alpha-1}$ are $\alpha$ ($\alpha\geq2$) neighboring and successive consecutive sequences with all their lengths odd, so that they are coupled in the sense of Proposition 3. Assuming that $T_{i-1}$ ($T_{i+\alpha}$) is not coupled to $T_i$ ($T_{i+\alpha-1}$), then the number of EMZMs associated with these $\alpha$ sequences is $\sum^{i+\alpha-1}_{j=i}v_j+\alpha$. Among these EMZMs, there are $\sum^{i+\alpha-1}_{j=i}v_j$ simple ones and $\alpha$ nontrivial ones. The wavefunctions for the latter spread over $2\sum^{i+\alpha-1}_{j=i}v_j+5$ lattice sites of the Majorana snake chain.
\par  \emph{Proof}: We first prove this Proposition for $\alpha=2$. We take $l_{u_i}=$ even since the case of odd $l_{u_i}$ can be analyzed similarly. By applying the equation set (\ref{EOMeoB}) to $T_i$ and $T_{i+1}$, it is easy to see that the two uncoupled sets of equations involving the $B$'s give $\frac{v_i+v_{i+1}}{2}+1$ simple EMZMs $d_{4j}$ and $d_{4j+2(v_i+1)}$, $j=\frac{l_{u_i}}{2},\frac{l_{u_i}}{2}+1,\cdots,\frac{l_{u_i}+v_i-1}{2}$. However, the other two sets of equations given by (\ref{EOMeoC}) are no longer independent since they share a common variable $A_{\frac{l_{u_i}+v_i+1}{2}}$. In fact, they are coupled together to form a single equation set consisting of $\frac{v_i+v_{i+1}}{2}+2$ equations, whose explicit form is listed in Appendix~\ref{AppC}.
If we eliminate the $C$'s from the equation set given by (\ref{succ}), we obtain a single equation for $A_{\frac{l_{u_i}}{2}}$, $A_{\frac{l_{u_i}+v_i+1}{2}}$, and $A_{\frac{l_{u_{i+1}}+v_{i+1}+1}{2}}$, see Eq.~(\ref{AAA}). The trivial solution of Eq.~(\ref{AAA}), i.e. $A_{\frac{l_{u_i}}{2}}=A_{\frac{l_{u_i}+v_i+1}{2}}=A_{\frac{l_{u_{i+1}}+v_{i+1}+1}{2}}=0$, leads to $\frac{v_{i}+v_{i+1}}{2}-1$ simple EMZMs $d_{2l_{u_i}+1},d_{2l_{u_i}+5},\cdots,d_{2l_{u_i}+2v_{i}-5}$ and $d_{2l_{u_{i+1}}+1},d_{2l_{u_{i+1}}+5},\cdots,d_{2l_{u_{i+1}}+2v_{i+1}-5}$. However, the rank-nullity theorem tells us that there are also two linearly independent nonzero solutions, which result in two nontrivial EMZMs. The wavefunctions of these two EMZMs spread over a segment of the snake chain of length $2v_i+2v_{i+1}+5$. As an illustration, we plot in Fig.~\ref{Fig2}(b) and (c) the wavefunctions for the two nontrivial EMZMs corresponding to $T_1=(8)$ and $T_2=(10,11,12)$ for a chain with $L=16$ sites. Thus, we have
\begin{eqnarray}
\left(\frac{v_i+v_{i+1}}{2}+1\right)+\left(\frac{v_{i}+v_{i+1}}{2}-1\right)+2=v_i+v_{i+1}+2\nonumber
\end{eqnarray}
EMZMs in total, among which there are $v_i+v_{i+1}$ simple ones, and two nontrivial ones.
\par The above arguments can be generalized to cases with $\alpha>2$. In general, if $T_i,T_{i+1},\cdots$, and $T_{i+\alpha-1}$ are $\alpha$ coupled successive consecutive sequences, then the $\alpha+1$ variables $A_{\frac{l_{u_i}}{2}},A_{\frac{l_{u_{i+1}}}{2}},\cdots,A_{\frac{l_{u_{i+\alpha-1}}}{2}}$, and $A_{\frac{l_{u_{i+\alpha-1}}+v_{i+\alpha-1}+1}{2}}$ will be coupled, yielding $\alpha$ linearly independent nontrivial solutions of $V$ besides the $\sum^{i+\alpha-1}_{j=i}v_j$ simple solutions. These $\alpha$ nontrivial EMZMs locate on a segment of length $2\sum^{i+\alpha-1}_{j=i}v_j+5$ (from site $2l_{u_i}-3$ to site $2l_{u_{i+\alpha-1}}+2v_{i+\alpha-1}-1$) of the Majorana snake chain. Proposition 4 is thus proved.
\par We are now in a position to prove our Theorem.
\par \emph{Proof of the Theorem}: Consider a generic set $\mathcal{S}=\{T_1,T_2,\cdots,T_n\}$. Two coupled (uncoupled) nearest-neighbor consecutive sequences $T_l$ and $T_{l+1}$ will be denoted as $T_l\sim T_{l+1}$ ($T_l\nsim T_{l+1}$). For any $j$, we further write $T_j\equiv T^{\mathrm{e}}_j$ or $T^{\mathrm{o}}_j$ to indicate the length of $T_j$ is even or odd. It is worth noting that the numbers of EMZMs associated with uncoupled strings of consecutive sequences can be counted \emph{independently}.
\par According to Proposition 3, any even sequence $T^{\mathrm{e}}_j$ is uncoupled with its neighboring consequences
\begin{eqnarray}
\cdots T^{\mathrm{e/o}}_{j-1}\nsim T^{\mathrm{e}}_j\nsim T^{\mathrm{e/o}}_{j+1}\cdots\nonumber
\end{eqnarray}
and the corresponding nubmer of EMZMs is just the length of $T^{\mathrm{e}}_j$, i.e.
\begin{eqnarray}
v_j=v_j+\xi(v_j),
\end{eqnarray}
since $v_j$ is an even number. Similarly, any $\alpha$ coupled consecutive sequences $T_i,T_{i+1},\cdots,T_{i+\alpha-1}$ appear in $\mathcal{S}$ in the following way
\begin{eqnarray}
\cdots T^{\mathrm{e/o}}_{i-1}\nsim T^{\mathrm{o}}_i\sim T^{\mathrm{o}}_{i+1}\sim \cdots\sim T^{\mathrm{o}}_{i+\alpha-1}\nsim T^{\mathrm{e/o}}_{i+\alpha}\cdots
\end{eqnarray}
and the corresponding number of EMZMs can be expressed as (according to Proposition 4)
\begin{eqnarray}
\sum^{i+\alpha-1}_{l=i}v_l+\alpha=\sum^{i+\alpha-1}_{l=i}\left[v_l+\xi(v_l)\right],
\end{eqnarray}
since all the $v_l$'s are odd. From Proposition 2, we note that the above equation is also valid for $\alpha=1$, i.e., a single odd sequence sandwiched by two neighboring sequences that are decoupled from it.
\par Taking into account all contributions from the even consecutive sequences, the coupled consecutive-sequence strings, and the single odd sequences sandwiched by two uncoupled sequences, the total number of EMZMs for $H^{\mathrm{(MF)}}_{\mathrm{OBC}}$ associated with $\mathcal{S}$ can be counted as
\begin{eqnarray}
\mathcal{N}=\sum^{n}_{i=1}\left[v_i+\xi(v_i)\right]=m+\sum^{n}_{i=1}\xi(v_i).
\end{eqnarray}
This completes the proof of the Theorem.
\section{Degeneracy of eigenstates in the transverse compass chain}\label{SecIV}
\subsection{The eigenstate-degeneracy problem}\label{examplesDG}
\par The high degeneracy of ground states of the compass model has been noticed early and is associated with the symmetry analysis~\cite{Dorier2005,Brzezicki07,You08}. It was soon recognized that any energy level in the whole energy spectra possesses the same degeneracy. Such macroscopic (usually exponential) degeneracy in compass systems arises from the existence of the so-called intermediate symmetries~\cite{Nussinov2015}, which are composed of a group of local symmetry operators~\cite{You10,Brzezicki13}. However, the highly degenerate ground-state manifold is vulnerable and can be completely lifted by an infinitesimal transverse field~\cite{Sun09,You14}. Recently it was realized the ground-state manifold is actually intact provided the magnetic field is absent at the odd sites but takes any finite value at the even sites~\cite{You2018}. In general, the presence of degenerate manifold necessitates the absence of fields on a fragment of the lattice sites. Below we present two examples to illustrate these observations.
\begin{figure}
\includegraphics[width=.52\textwidth]{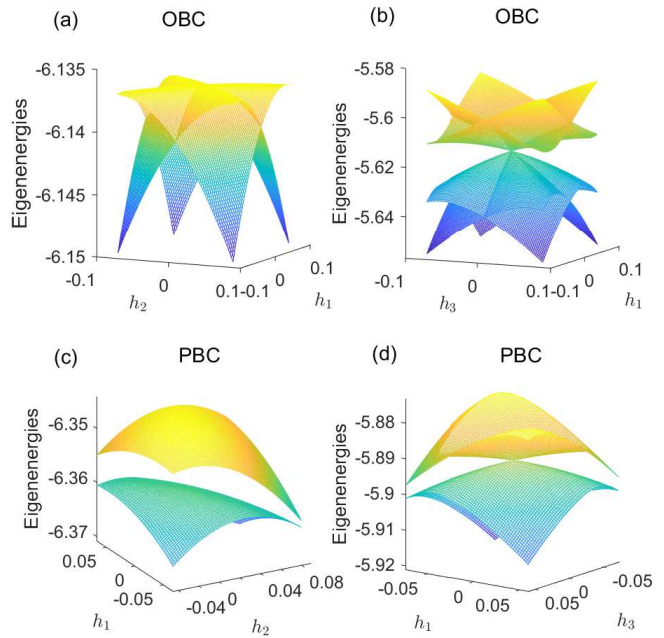}
\caption{Evolution of low-lying eigenenergies of $H_{\mathrm{OBC}}$ [(a) and (b)] and $H_{\mathrm{PBC}}$ [(c) and (d)] when the magnetic fields on two certain sites are varied. Calculations are performed for $L=8$, $J^x_j=1$, and $J^y_j=0.8$ ($j=1,2,\cdots,L/2$). In each panel, the remaining $L-2$ fixed magnetic fields are chosen to be $h_l=1$ ($0.2$) for odd (even) $l$.}
\label{Fig3}
\end{figure}
\par We plot in the four panels of Fig.~\ref{Fig3} the lowest several eigenenergies of $H_{\mathrm{OBC}}$ or $H_{\mathrm{PBC}}$ when the magnetic field on two of the lattice sites are varied. Figure~\ref{Fig3}(a) shows the energies of the ground state and first excited state of $H_{\mathrm{OBC}}$ as functions of $h_1$ and $h_2$. It can be seen that level crossing occurs on the two lines with $h_1=0$ and $h_2=0$, indicating that the ground state is two-fold degenerate if one of $h_1$ and $h_2$ vanishes. Note that no additional degeneracy is introduced if both $h_1$ and $h_2$ are set zero. However, if we introduce a boundary term to form a periodic chain, the ground state becomes nondegenerate for arbitrary $h_1$ and $h_2$ [Fig.~\ref{Fig3}(c)].
\par The results for varying $h_1$ and $h_3$ are presented in Fig.~\ref{Fig3}(b) and (d). In the case of the OBC, the level crossing still exists along the lines $h_1=0$ and $h_3=0$. However, we observe from Fig.~\ref{Fig3}(b) that the lowest four energy levels touch at the point $(h_1,h_3)=(0,0)$, which indicates that the ground state of $H_{\mathrm{OBC}}$ is four-fold degenerate at this point. For the case of PBC, the level crossing for $h_1=0, h_3\neq 0$ and $h_1\neq 0, h_3=0$ disappears, and the four-fold degeneracy is reduced to a two-fold degeneracy at $(h_1,h_3)=(0,0)$ [Fig.~\ref{Fig3}(d)].
\par We see from the above examples that the properties of the eigenstate-degeneracy in the transverse compass chain depend not only on the details of the vanishing fields, but also on the boundary conditions imposed. We will see in the next subsection that all these observations can be clearly demonstrated in the framework of the emergent EMZMs.
\subsection{Interpretation in terms of the EMZMs}
\par Previously, the ground-state degeneracy of the compass model described by $H_{\mathrm{OBC}}$ and $H_{\mathrm{PBC}}$ has been discussed with the help of symmetry operators that commute with the Hamiltonian~\cite{Brzezicki07,You08,You10,Brzezicki13,You14}. As an example, consider an open compass chain in the absence of the transverse field. Then it is easy to check that the following bond operators~\cite{You14}
\begin{eqnarray}
\label{opx}
X_{2j}&=& \sigma^x_{2j} \sigma^x_{2j+1},  \quad
Y_{2j-1}= \sigma^y_{2j-1} \sigma^y_{2j},
\end{eqnarray}
possess local $Z_2$ symmetries and commute with the Hamiltonian. In addition, the two operators on nearest-neighboring bonds anticommute with each other, i.e., $\{Y_{2j-1},X_{2j}\}=0$. Let $|\Phi_{\mathrm{OBC}}\rangle$ be a common eigenstate of $H_{\mathrm{OBC}}$ and the tuple $\vec{Y}\equiv (Y_1,Y_3,\cdots,Y_{\frac{L}{2}-1})$ with eigenvalues $E$ and $(y_1,y_3,\cdots,y_{\frac{L}{2}-1})$, respectively
\begin{eqnarray}
\label{opx}
H_{\mathrm{OBC}}|\Phi_{\mathrm{OBC}}\rangle&=& E|\Phi_{\mathrm{OBC}}\rangle,\nonumber\\
\vec{Y}|\Phi_{\mathrm{OBC}}\rangle&=&(y_1,y_3,\cdots,y_{\frac{L}{2}-1})|\Phi_{\mathrm{OBC}}\rangle,
\end{eqnarray}
where $y_{2j-1}=\pm 1$ since $Y^2_{2j-1}=1$. Now consider the state $X_{2j}|\Phi_{\mathrm{OBC}}\rangle$, which is another eigenstate of $H_{\mathrm{OBC}}$ with the same energy $E$ according to the relation $[H_{\mathrm{OBC}},X_{2j}]=0$. Then one can be convinced that the two states $|\Phi_{\mathrm{OBC}}\rangle$ and $X_{2j}|\Phi_{\mathrm{OBC}}\rangle$ are distinct since the latter is an eigenstate of $Y_{2j-1}$ with eigenvalue $-y_{2j-1}$ by considering $\{X_{2j}, Y_{2j-1}\}=0$. We can apply the above procedure repeatedly until one exhausts all the $2^{\frac{L}{2}}$ eigenstates in the degenerate manifold containing $|\Phi_{\mathrm{OBC}}\rangle$.
\par In principle, there always exists a complete set of symmetry operators that can be used to interpret the underlying degeneracy of eigenstates. However, the construction of these local symmetry operators becomes intricate for general configurations of the vanishing inhomogeneous fields. A complete and compact form of all the independent symmetry operators is usually difficult to identify unless in specific cases. Below we will see that the eigenstate-degeneracy and level crossing observed in the last subsection can be most clearly demonstrated in terms of the EMZMs. Furthermore, the explicit forms of the wavefunctions for the EMZMs provide a straightforward means to construct the symmetry operators in the spin representation. 

\par Suppose there are $\mathcal{N}$ EMZMs determined by some given set $\mathcal{S}$. Since the $2L$ eigenvalues of $\mathcal{H}_{\mathrm{OBC}}$ always appear in pairs, the total number of the corresponding fermionic zero modes is thus $\mathcal{N}/2$ [cf. Eq.~(\ref{HOBCff})]. As a result, the $2^L$ eigenstates of the fermion model $H^{\mathrm{(F)}}_{\mathrm{OBC}}$ given by Eq.~(\ref{HOBCff}), and hence of the spin model $H_{\mathrm{OBC}}$, are divided into $2^{L-\frac{\mathcal{N}}{2}}$ equivalence classes, each of which form a degenerate manifold possessing a definite energy and having $2^{\frac{\mathcal{N}}{2}}$ degenerate eigenstates. Note that only half of the $2^{\frac{\mathcal{N}}{2}}$ degenerate states in each manifold have even fermion parity, since for the OBC both the even and odd numbers of excitations of the fermionic zero modes are physically allowed.
\par For the case of the PBC in the spin representation, our foregoing analysis in the Majorana space is still valid. In practice, due to the presence of the boundary term, $H_{\mathrm{PBC}}$ can be written in the Majorana representation as~\cite{PRB2018}
\begin{eqnarray}
H^{\mathrm{(MF)}}_{\mathrm{PBC}}=\sum_{\sigma=\pm1}\frac{1+\sigma P}{2}H^{\mathrm{(MF)}}_{\mathrm{PBC},\sigma} \frac{1+\sigma P}{2},
\end{eqnarray}
where
\begin{eqnarray}
H^{\mathrm{(MF)}}_{\mathrm{PBC},\sigma}=\frac{i}{4}\mathcal{D}^T\mathcal{H}_{\mathrm{PBC},\sigma}\mathcal{D},
\end{eqnarray}
and $P$ is the fermion parity operator. The index $\sigma=+1$ ($-1$) indicates the subspace with even (odd) numbers of fermionic excitations. 
The $2L\times2L$ matrix $\mathcal{H}_{\mathrm{PBC},\sigma}$ is similar to $\mathcal{H}_{\mathrm{OBC}}$, but with the boundary elements
\begin{eqnarray}
\mathcal{H}_{\mathrm{PBC},\sigma}(2,2L-1)=-\mathcal{H}_{\mathrm{PBC},\sigma}(2L-1,2)=-\sigma J^y_{\frac{L}{2}}\nonumber
\end{eqnarray}
included. Our analysis based on the coupled eigen-equations still applies to $\mathcal{H}_{\mathrm{PBC},\sigma}$, with the understanding that $(\cdots,L,1,\cdots)$ should be considered as a consecutive sequence if $h_L=h_1=0$. Any eigenstate $|\psi_{\mathrm{PBC}}\rangle$ of $H^{\mathrm{(MF)}}_{\mathrm{PBC}}$ has a definite fermion parity $\sigma$ (though the practical determination of $\sigma$ for the ground state is not straightforward). Provided that there are no eigenstates in the subspace labelled by $-\sigma$ having the same energy as that of $|\psi_{\mathrm{PBC}}\rangle$, the degenerate manifold containing $|\psi_{\mathrm{PBC}}\rangle$ must belong to the $\sigma$-subspace, since only even numbers of excitations in the $\sigma$-subspace are physically allowed, which reduces the number of states in each degenerate manifold from $2^{\frac{\mathcal{N}}{2}}$ to $2^{\frac{\mathcal{N}}{2}-1}$~\cite{OEdeg}. It is easy to check that the ground-state degeneracy found in Fig.~\ref{Fig3} can be explained in terms of the emergent EMZMs.
\par As an interesting example, let us consider the case in which all fields on the odd sublattice are turned off. In this case we have $\mathcal{S}=\{T_1,T_2,\cdots,T_{\frac{L}{2}}\}$, where $T_j=(2j-1),~\forall j$. Thus, any nearest-neighboring consecutive sequences $T_i$ and $T_{i+1}$ are successive and coupled. According to Proposition 4, there are $\frac{L}{2}$ simple EMZMs and $\frac{L}{2}$ nontrivial EMZMs. We conclude that each eigenstate of $H_{\mathrm{OBC}}$ ($H_{\mathrm{PBC}}$) is $2^\frac{L}{2}$-fold ($2^{\frac{L}{2}-1}$-fold) degenerate, consistent with the qualitative arguments based on the symmetry operators~\cite{You2018}.
\par Since the EMZMs and their multiple products all commute with the Hamiltonian, the obtained explicit forms of the EMZMs provide an easy way to find out the symmetry operators in the spin representation by utilizing the reverse transformations of Eq.~(\ref{JW}) and Eq.~(\ref{MF}). For example, consider the case of $h_1=h_2=0$, then the two isolated EMZMs $d_1$ and $d_4$ can form two conserved symmetry operators $d_1=\sigma^x_1$ and $id_1d_4=\sigma^y_1\sigma^y_2$. It is easy to see that $[H_{\mathrm{OBC}},\sigma^x_1]=[H_{\mathrm{OBC}},\sigma^y_1\sigma^y_2]=0$ and $\{\sigma^x_1,\sigma^y_1\sigma^y_2\}=0$, implying that any eigenstate is two-fold degenerate for $h_1=h_2=0$ [Fig.~\ref{Fig3} (a)]. For $h_1=h_3=0$, the two isolated EMZMs $d_1$ and $d_5$ can form two independent symmetry operators $d_1=\sigma^x_1$ and $id_1d_5=\sigma^y_1\sigma^z_2\sigma^x_3$~\cite{You2018}. The remaining two nontrivial EMZMs will generate another pair of symmetry operators whose forms could be more complicated. 

\section{Exact solution for alternating nearest-neighbor interactions and staggered magnetic fields}\label{SecV}
\par After finding out the EMZMs according to the partition properties of the set $\mathcal{S}$, let us see whether $H_{\mathrm{OBC}}$ can host almost Majorana zero modes. In principle, this depends on the specific choice of the fields $\{h_j\}$, and it is not easy to tell whether the AMZMs exist for general configurations the inhomogeneous model. However, this question can be answered rigorously in the special case of an open compass chain with alternating nearest-neighbor interactions and under an alternating field, for which exact solutions exist. In this section, we set $J^x_j=J_x$, $J^y_j=J_y$, and $h_{2j-1}=\mu_1$, $h_{2j}=\mu_2$ ($j=1,2,\cdots,L/2$), and assume
\begin{eqnarray}
\gamma\equiv J_y/J_x>0,
\end{eqnarray}
since the other cases can be reached by appropriate unitary transformations. The Hamiltonian $H_{\mathrm{OBC}}$ given by Eq.~(1) becomes
\begin{eqnarray}\label{Hobcalt}
H^{\mathrm{(alt)}}_{\mathrm{OBC}}&=&-J_x\sum^{\frac{L}{2}}_{j=1}\sigma^x_{2j-1}\sigma^x_{2j}-J_y\sum^{\frac{L}{2}-1}_{j=1}\sigma^y_{2j}\sigma^y_{2j+1} \nonumber\\
&&-\mu_1\sum^{\frac{L}{2}}_{j=1} \sigma^z_{2j-1}-\mu_2\sum^{\frac{L}{2}}_{j=1}  \sigma^z_{2j}.
\end{eqnarray}
\subsection{$\mu_1\neq0$ and $\mu_2\neq 0$}\label{SecVA}
\par From Proposition 1, there are no EMZMs for $\mu_1,\mu_2\neq0$. To obtain the $2L$ nonzero eigenvalues of $\mathcal{H}_{\mathrm{OBC}}$, we employ a plane-wave-like \emph{ansatz}~\cite{PRB2017,Grim}:
\begin{eqnarray}
A_j&=&R_a e^{ik(4j-3)}+S_a e^{-ik(4j-3)},\nonumber\\
B_j&=&R_b e^{ik(4j-2)}+S_b e^{-ik(4j-2)},\nonumber\\
C_j&=&R_c e^{ik(4j-1)}+S_c e^{-ik(4j-1)},\nonumber\\
D_j&=&R_d e^{ik(4j)}+S_d e^{-ik(4j)},
\end{eqnarray}
with $j=1,2,\cdots,\frac{L}{2}$. Here, the $R$'s and $S$'s are $j$-independent coefficients to be determined. The form of the above ansatz is inspired by the quasi-periodic structure of $H^{\mathrm{(alt)}}_{\mathrm{OBC}}$.
\par We substitute the above ansatz in the bulk equations given by Eqs.~(\ref{bulk0_1}) to (\ref{bulk0_4}), and compare the coefficients of the same power of $e^{\pm ikj}$ on both sides to obtain
\begin{eqnarray}
\label{bulkRa}
-\mu_{1}  R_b e^{ik}&=&\lambda R_a,\\
\label{bulkRd}
\mu_{2} R_c e^{-ik}&=&\lambda R_d,\\
\label{bulkRc}
-J_x R_b e^{-i2k}-\mu_{2} R_d  -J_y R_b e^{i2k}&=&\lambda R_c e^{-ik},\\
\label{bulkRb}
J_y R_c e^{-i2k}+\mu_{1} R_a  +J_x R_c e^{i2k }&=&\lambda R_b e^{ik },
\end{eqnarray}
and
\begin{eqnarray}
\label{bulkSa}
-\mu_{1}  S_b e^{-ik}&=&\lambda S_a,\\
\label{bulkSd}
\mu_{2} S_c e^{ ik}&=&\lambda S_d,\\
\label{bulkSc}
-J_x S_b e^{ i2k}-\mu_{2} S_d  -J_y S_b e^{-i2k}&=&\lambda S_c e^{ ik},\\
\label{bulkSb}
J_y S_c e^{ i2k}+\mu_{1} S_a  +J_x S_c e^{-i2k }&=&\lambda S_b e^{-ik }.
\end{eqnarray}
We can eliminate $R_a$, $R_d$, $S_a$, and $S_d$ by using the first two equations in the last expressions to get
\begin{eqnarray}\label{bulkcond1red}
\left(
  \begin{array}{cccc}
    \lambda J_x  ( e^{-i2k} +\gamma e^{i2k}) & (\lambda^2+\mu^2_2)   e^{-ik}  \\
   (\lambda^2+\mu^2_1)   e^{ik } & -\lambda  J_x (\gamma e^{-i2k} +   e^{i2k })  \\
  \end{array}
\right)\left(
         \begin{array}{c}
           R_b \\
           R_c \\
         \end{array}
       \right)=0,\nonumber\\
\end{eqnarray}
and
\begin{eqnarray}\label{bulkcond2red}
\left(
  \begin{array}{cccc}
    \lambda  J_x ( e^{i2k} +\gamma e^{-i2k}) & (\lambda^2+\mu^2_2)   e^{ik}  \\
   (\lambda^2+\mu^2_1)   e^{-ik } & -\lambda  J_x (\gamma e^{i2k} + e^{-i2k })  \\
  \end{array}
\right)\left(
         \begin{array}{c}
           S_b \\
           S_c \\
         \end{array}
       \right)=0.\nonumber\\
\end{eqnarray}
To obtain nontrivial solutions of $(R_b, R_c)$ and $(S_b,S_c)$, the determinants of the $2\times2$ matrices appearing in the above matrix equations
\begin{eqnarray}\label{detl4}
\det=-[\lambda^4+(\delta_k+\mu^2_1+\mu^2_2)\lambda^2+\mu^2_1 \mu^2_2],
\end{eqnarray}
must be zero, yielding four branches of \emph{nonzero} single-particle dispersions
\begin{eqnarray}\label{dispersion}
\lambda^{(\pm)}_{k,\pm}=\pm\frac{i}{\sqrt{2}} \sqrt{(\delta_k+\nu_+)\pm\sqrt{\delta_k(\delta_k+2\nu_+)+\nu_-^2}},
\end{eqnarray}
where
\begin{eqnarray}\label{deltak}
\delta_k&\equiv&J^2_x(1+\gamma^2+2\gamma\cos 4k),\nonumber\\
\nu_{\pm}&\equiv&\mu^2_1\pm\mu^2_2,
\end{eqnarray}
and the superscript $(\pm)$ denotes the overall sign of $\lambda$.
\par \par We still need to determine the allowed values of $k$. To this end, we have to apply the ansatz in the two boundary equations given by Eq.~(\ref{boud0_1}) and Eq.~(\ref{boud0_2}). By using the bulk equations Eqs.~(\ref{bulkRa})-(\ref{bulkSb}) to eliminate $R_a$, $R_d$, $S_a$, and $S_d$ in the boundary equations, and using the fact that $\lambda\neq0$, we obtain the following equations for $(R_b, S_b)$ and $(R_c,S_c)$
\begin{eqnarray}\label{bound1final}
&&0= R_b e^{ik(2L+2)}   + S_b e^{-ik(2L+2)} ,
\end{eqnarray}
and
\begin{eqnarray}\label{bound2final}
&& 0=R_c    e^{-i k}   +S_c e^{ ik}.
\end{eqnarray}
By adding up the first equation of (\ref{bulkcond1red}) and the first equation of (\ref{bulkcond2red}), we get
\begin{eqnarray} \label{addup}
  R_b (  e^{-i2k} +\gamma e^{i2k})+S_b ( e^{i2k} +\gamma e^{-i2k})&=&0,
\end{eqnarray}
where we have used Eq.~(\ref{bound2final}).
\par Now we combine Eq.~(\ref{bound1final}) and Eq.~(\ref{addup}) to obtain
\begin{eqnarray} \label{detk}
\left(
  \begin{array}{cc}
    e^{ik(2L+2)} & e^{-ik(2L+2)} \\
  e^{-i2k} +\gamma e^{i2k} &  e^{i2k} +\gamma e^{-i2k} \\
  \end{array}
\right)\left(
         \begin{array}{c}
           R_b \\
           S_b \\
         \end{array}
       \right)=0.
\end{eqnarray}
Letting the determinant of the above matrix be zero, we finally obtain the desired quantization condition that determines the allowed values of the  wavenumber $k$
\begin{eqnarray}
\label{detkfinal}
\frac{\sin 2k(L+2)}{\sin 2kL}&=&-\gamma.
\end{eqnarray}
It is intriguing to note that the allowed $k$'s do \emph{not} depend on the fields $\mu_1$ and $\mu_2$.
\par It can be shown that (see Appendix~\ref{AppB}) we only need to find out the solutions on the interval $k\in[0,\frac{\pi}{4}]$: For $\gamma<\frac{L+2}{L}$, there are $\frac{L}{2}$ real solutions; For $\gamma>\frac{L+2}{L}$, there exist $\frac{L}{2}-1$ real solutions, and a single complex solution of the form~\cite{Lieb,finiteIsing}
\begin{eqnarray} \label{complexk}
\tilde{k}=\frac{\pi}{4}+ i\tilde{k}_1,
\end{eqnarray}
where $\tilde{k}_1$ is the solution of the equation
\begin{eqnarray} \label{sinhsinh}
\frac{\sinh 2k(L+2)}{\sinh 2kL} =\gamma
\end{eqnarray}
on the interval $[0,\infty)$. It is useful to note that $\frac{\sinh 2k(L+2)}{\sinh 2kL}\to e^{4k}$ in the thermodynamic limit $L\to\infty$, so that
\begin{eqnarray} \label{k1ln}
\tilde{k}_1\to \frac{1}{4}\ln\gamma,~~L\to\infty.
\end{eqnarray}
The four eigenvalues corresponding to the complex solution $\tilde{k}$ are still given by Eq.~(\ref{dispersion}), but with $\delta_k$ replaced by
\begin{eqnarray}\label{deltaktilde}
\tilde{\delta}_{\tilde{k}}=J^2_x(1+\gamma^2-2\gamma\cosh 4\tilde{k}_1).
\end{eqnarray}
We can show that $\delta_k$ and $\tilde{\delta}_{\tilde{k}}$ are both nonnegative (Appendix~\ref{AppB}). Since $\mu_1\neq 0$ and $\mu_2\neq 0$, we see from Eq.~(\ref{dispersion}) that the eigenvalues $\lambda^{(\pm)}_{k,\pm}$ and $\lambda^{(\pm)}_{\tilde{k},\pm}$ never approach zero even in the thermodynamic limit $L\to\infty$. As a result, there are no AMZMs for $\mu_1\neq 0$ and $\mu_2\neq 0$. In other words, $H_{\mathrm{OBC}}$ does not have a quantum phase transition.
\par As a byproduct, we can also write down the eigenvectors of $\mathcal{H}_{\mathrm{OBC}}$:
\par 1) For $\gamma<\frac{L+2}{L}$, the eigenvector $V^{(\pm)}_{k,\pm}$ corresponding to $\lambda^{(\pm)}_{k,\pm}$ is
\begin{eqnarray}\label{Vecreal}
&&V^{(\pm)}_{k,\pm}=N_{k,\pm}\left(A^{(\pm)}_{1,\pm},B^{(\pm)}_{1,\pm},\cdots,C^{(\pm)}_{\frac{L}{2},\pm},D^{(\pm)}_{\frac{L}{2},\pm}\right)^T,
\end{eqnarray}
where
\begin{eqnarray}\label{Vecreal1}
A^{(\pm)}_{j,\pm}&=& \frac{  \mu_1 J_x[\gamma\sin 4k(j-1) +\sin 4kj]  }{ \lambda^{(\pm)2}_{k,\pm}+\mu^2_1 },\nonumber\\
B^{(\pm)}_{j,\pm}&=& \frac{-\lambda^{(\pm)}_{k,\pm} J_x [ \gamma\sin 4k(j-1)+ \sin 4kj]}{ \lambda^{(\pm)2}_{k,\pm}+\mu^2_1 },\nonumber\\
C^{(\pm)}_{j,\pm}&=& -\sin4kj,\nonumber\\
D^{(\pm)}_{j,\pm}&=&-\frac{\mu_2  }{\lambda^{(\pm)}_{k,\pm}}\sin 4kj,
\end{eqnarray}
and $N_{k,\pm}$ are two \emph{real} normalization factors. Since $\lambda^{(+)}_{k,\pm}=\lambda^{(-)*}_{k,\pm}$, so the reality of $N_{k,\pm}$ ensures
\begin{eqnarray}
V^{(+)}_{k,\pm}=V^{(-)*}_{k,\pm},
\end{eqnarray}
as required by Eq.~(\ref{VVstar}).
\par 2) For $\gamma>\frac{L+2}{L}$, the eigenvectors for the corresponding $\frac{L}{2}-1$ real solutions of $k$ are also given by Eq.~(\ref{Vecreal}) and (\ref{Vecreal1}). For the complex solution $\tilde{k}$, the eigenvector corresponding to $\lambda^{(\pm)}_{\tilde{k},\pm}$ reads [use the relation $\sin (ix)=i\sinh x$]
\begin{eqnarray}\label{Veccomp}
&&V^{(\pm)}_{\tilde{k},\pm}=\tilde{N}_{\tilde{k},\pm}\left(\tilde{A}^{(\pm)}_{1,\pm},\tilde{B}^{(\pm)}_{1,\pm},\cdots,\tilde{C}^{(\pm)}_{\frac{L}{2},\pm},\tilde{D}^{(\pm)}_{\frac{L}{2},\pm}\right)^T,
\end{eqnarray}
where
\begin{eqnarray}\label{Veccomp1}
\tilde{A}^{(\pm)}_{j,\pm}&=&(-1)^{j-1} \frac{  \mu_1 J_x[\gamma\sinh 4\tilde{k}_1(j-1) -\sinh 4\tilde{k}_1j]  }{ \lambda^{(\pm)2}_{\tilde{k},\pm}+\mu^2_1 },\nonumber\\
\tilde{B}^{(\pm)}_{j,\pm}&=&(-1)^j \frac{\lambda^{(\pm)}_{\tilde{k},\pm}J_x  [\gamma\sinh 4\tilde{k}_1(j-1) -\sinh 4\tilde{k}_1j]}{ \lambda^{(\pm)2}_{\tilde{k},\pm}+\mu^2_1 },\nonumber\\
\tilde{C}^{(\pm)}_{j,\pm}&=& (-1)^{j-1} \sinh4\tilde{k}_1j,\nonumber\\
\tilde{D}^{(\pm)}_{j,\pm}&=& (-1)^{j-1}\frac{\mu_2  }{\lambda^{(\pm)}_{\tilde{k},\pm}}\sinh 4\tilde{k}_1j,
\end{eqnarray}
and $\tilde{N}_{\tilde{k},\pm}$ are another two real normalization factors.
\subsection{$\mu_1=0$ and $\mu_2\neq 0$}\label{SecB}
\par If the fields on the odd sublattice are turned off, i.e., $\mu_1=0$ and $\mu_2\neq0$, then there are $L$ EMZMs according the Theorem. The remaining $L$ nonzero modes can also be obtained using the ansatz method, yielding two branches of dispersion
\begin{eqnarray}\label{dispersionB}
\lambda^{(\pm)}_{k }=\pm i\sqrt{\delta_k+\mu^2_2}.
\end{eqnarray}
Thus, there are still no AMZMs except for the $\frac{L}{2}$ exact ones mentioned above.
\subsection{$\mu_1=0$ and $\mu_2=0$}\label{SecC}
\par For $\mu_1=\mu_2=0$, the Hamiltonian given by Eq.~(\ref{Hobcih}) reduces to the one-dimensional limit of the Kitaev honeycomb model~\cite{Feng07,Kitaev2006}. It is known that the compass chain in the absence of external fields can be mapped onto the transverse Ising chain on the dual lattice through a spin duality transformation~\cite{Feng07}. Therefore, an Ising-type quantum phase transition is expected to occur at $\gamma=1$ in the thermodynamic limit. Actually, the order/disordered phase of the transverse Ising chain on the dual lattice correspond to the presence of two hidden string order parameters in the original space~\cite{Feng07}.
\par The mapping between the two models can also be revealed in the Majorana representation by relabelling the MFs in the second row of the comb (the first row of the comb is completely decoupled). Let
\begin{eqnarray}
d'_{2j-1}=d_{4j-2},~~d'_{2j}=d_{4j-1},~~j=1,2,\cdots,\frac{L}{2},
\end{eqnarray}
then the Hamiltonian of the snake chain given by Eq.~(\ref{HMFDD}) in the absence of the magnetic field becomes
\begin{eqnarray}
H^{(\mathrm{MF})}_{\mathrm{OBC}}&=&i\sum^{\frac{L}{2}}_{j=1} J^x_{j}d'_{2j-1}d'_{2j} -i\sum^{\frac{L}{2}-1}_{j=1} J^y_{j}d'_{2j}d'_{2j+1}.
\end{eqnarray}
If we define ordinary fermions via
\begin{eqnarray}
\left(
  \begin{array}{c}
     d'_{2j-1} \\
     d'_{2j} \\
  \end{array}
\right)=\Gamma_2 \left(
  \begin{array}{c}
   a'_j \\
   a'^\dag_j \\
  \end{array}
\right),~j=1,2,\cdots,\frac{L}{2}
\end{eqnarray}
then $H^{(\mathrm{MF})}_{\mathrm{OBC}}$ can be written as
\begin{eqnarray}\label{Kitaevpwave}
H^{(\mathrm{KT})}_{\mathrm{OBC}}&=&\sum^{\frac{L}{2}-1}_{j=1} J^y_{j}(a'^\dag_ja'_{j+1}+a'_{j+1}a'_j+\mathrm{H.c.})\nonumber\\
&&+\sum^{\frac{L}{2}}_{j=1} 2J^x_{j}\left(a'^\dag_j a'_j-\frac{1}{2}\right),
\end{eqnarray}
which is nothing but an inhomogeneous open Kitaev $p$-wave superconducting chain of length $\frac{L}{2}$ with equal hopping and pairing strengthes. By further performing the inverse Jordan-Wigner transformation to the above equation, we achieve the mapping from the compass chain to an inhomogeneous transverse Ising chain.
\par We now study the full spectrum of the open compass chain $H^{\mathrm{(alt)}}_{\mathrm{OBC}}(\mu_1=0,\mu_2=0)$ using the results obtained in Sec.~\ref{SecVA}. According to the Theorem, there are still $L$ EMZMs, corresponding to the $L$ isolated MFs in the first row of the comb. From Eq.~(\ref{dispersionB}), the corresponding dispersion in this case becomes
\begin{eqnarray}
\lambda^{(\pm)}_{k }=\pm i\sqrt{\delta_k}.
\end{eqnarray}
We focus on the case of $\gamma>\frac{L+2}{L}$, for which $\lambda^{\mathrm{(\pm)}}_{k}$ for the real solutions of $k$ are always finite. The dispersion $\lambda^{\mathrm{(\pm)}}_{\tilde{k}}$ for the complex solution $\tilde{k}$ can be reexpressed with the help of Eq.~(\ref{deltaktilde}) as
\begin{eqnarray}
\lambda^{\mathrm{(\pm)}}_{\tilde{k}}&=&\pm 2i J_x\sinh 4\tilde{k}_1\sqrt{\frac{ e^{-4\tilde{k}_1}}{e^{4\tilde{k}_1L}-1}\left(\frac{ e^{-4\tilde{k}_1}}{e^{4\tilde{k}_1L}-1}+ \gamma \right)}.\nonumber\\
\end{eqnarray}
In the thermodynamic limit $L\to \infty$, we have $e^{4\tilde{k}_1L}\gg 1$ and $\gamma\to e^{4\tilde{k}_1}$, so that $\lambda^{\mathrm{(\pm)}}_{\tilde{k}}$  approaches zero \emph{exponentially}
\begin{eqnarray}
\lambda^{\mathrm{(\pm)}}_{\tilde{k}}&=&\pm 2i J_x\sinh 4\tilde{k}_1 e^{-2\tilde{k}_1L}.
\end{eqnarray}
Thus, the onset of the complex solution $\tilde{k}$ for $\gamma>\frac{L+2}{L}$ indicate two AMZMs in the thermodynamic limit. From Eq.~(\ref{Veccomp1}), the corresponding eigenvectors $V^{(\pm)}_{\tilde{k} }$ for finite $L$ have components ($j=1,2,\cdots,\frac{L}{2}$)
\begin{eqnarray}\label{Veccomp1_1}
\tilde{A}^{(\pm)}_j&=&\tilde{D}^{(\pm)}_j=0,\nonumber\\
\tilde{B}^{(\pm)}_j&=&(-1)^j \frac{  J_x [\gamma\sinh 4\tilde{k}_1(j-1) -\sinh 4\tilde{k}_1j ]}{\lambda^{(\pm) }_{\tilde{k}}},\nonumber\\
\tilde{C}^{(\pm)}_j&=& (-1)^{j-1} \sinh4\tilde{k}_1j.
\end{eqnarray}
Note that $\tilde{B}^{(+)}_j=-\tilde{B}^{(-)}_j$ and $\tilde{C}^{(+)}_j=\tilde{C}^{(-)}_j$. According to Eq.~(\ref{getW}), we can in turn obtain the wavefunctions for the two AMZMs [we arrange $\lambda^{\mathrm{(\pm)}}_{\tilde{k}}$ to be the last pair of eigenvalues in Eq.~(\ref{Theta})]
\begin{eqnarray}\label{wfAMZM}
W_{{2L-1},:}&=&\frac{\tilde{N}_{\tilde{k}}}{\sqrt{2}}\left(0,0,\tilde{C}^{(+)}_1,0,\cdots,0,0,\tilde{C}^{(+)}_{\frac{L}{2}},0\right),\nonumber\\
W_{{2L},:}&=&i\frac{\tilde{N}_{\tilde{k}}}{\sqrt{2}}\left(0,\tilde{B}^{(+)}_1,0,0,\cdots,0,\tilde{B}^{(+)}_{\frac{L}{2}},0,0\right).\nonumber\\
\end{eqnarray}
In the thermodynamic limit, we have
\begin{eqnarray}\label{wfAMZM}
\tilde{B}^{(\pm)}_j&\to&\pm \frac{i}{2}(-1)^j\gamma^{\frac{L}{2}-j+1},
\end{eqnarray}
which shows that the two AMZMs $d'_{L}$ and $d''_{L}$ are indeed Majorana edge zero modes locating at the right and left ends of the lower leg of the snake chain, respectively.
\par
\section{Conclusions and Discussions}\label{SecVI}
\par In this work, we have studied the emergence of exact Majorana zero modes in interacting spin chains. We focus on a one-dimensional quantum compass model with both the nearest-neighbor interactions and transverse fields varying over space. To diagonalize the model, we first perform a Jordan-Wigner transformation to map the spin model into a fermionic one. We then define a set of Majorana operators by linearly combining the fermion creation/annihilation operators on individual sites. The finally obtained quadratic Majorana-fermion model provides us a convenient representation to study the existence of EMZMs.
\par Working in the Majorana fermion representation, we are able to derive an analytical expression for the number of the emergent EMZMs and find that it depends on how the lattice sites on which the magnetic fields vanish are divided into consecutive sequences. As byproducts, we also derive explicit forms of the EMZMs. It is found that highly nonlocal EMZMs can occur if the fields vanish at lattice sites corresponding to several coupled consecutive sequences of odd lengths. These exact results on the EMZMs are then employed to completely explain the intriguing dependence of the eigenstate-degeneracy on the inhomogeneous fields observed in prior literatures. We finally concentrate on a special case of the model with alternating nearest-neighbor interactions and staggered transverse fields. Due to the quasi-periodic structure of the model in the bulk, we propose a plane-wave ansatz to exactly solve the model, and obtain the analytical forms of the single-particle dispersions and the eigenstates in finite systems. We show that two AMZMs emerge if the fields on both the two sublattices are turned off.

\par Finally, we would like to discuss possible experimental platforms that can realize our model and the control of the transverse fields.  Considering that the quantum Ising chain in the presence of both a transverse and a longitudinal magnetic field has recently been simulated using the nuclear magnetic resonance quantum simulator~\cite{PRX2017}, such a setup may also serve as a suitable simulator to realize our model. Another possible setup to realize the current model is the superconducting quantum circuits~\cite{Youjq,You2018}. In this scheme, the compass chain given by Eq.~(\ref{Hobcih}) can be built with superconducting charge qubits  placed at each site, each of which is composed of a direct current superconducting quantum interference device (dc SQUID) with two identical Josephson junctions. The magnetic flux threading the individual SQUID can be used to control the effective magnetic field~\cite{Youjq}. Recently, the spin-1/2 XXZ Heisenberg chain in a transverse field has been realized using the technique of nanoscale-arrangement of magnetic atoms~\cite{natphy2016}, which provides another candidate to simulate the transverse compass model. In practice, the local fields exerted on the quasi-one-dimensional magnetic materials can be induced by the proximity effect of a ferromagnet~\cite{Zvyagin2013}, and the strength of the field can be tuned by varying the distance between the spin chain and the ferromagnet. The extinguishing of local fields can be realized by substitution of nonmagnetic ions for magnetic ones. To detect the EMZMs, one can detect the level crossings through the differential conductance at each site measured by real-space sensitive spectroscopies (e.g., scanning tunnelling microscopy) when the transverse fields are varied~\cite{natphy2016,Mila2016}.
\par We emphasize that the advantage of the system considered here lies in its exact solvability and mathematical rigor. The exact results for the wavefunctions of the EMZMs and AMZMs may serve as a basis for upcoming studies of quench dynamics and braiding of Majorana modes in related systems.
\begin{acknowledgments}
We thank Wojciech Brzezicki for insightful discussions on the symmetry operators.  This work was supported by the NSFC under Grant No. 11705007, and partially by the Beijing Institute of Technology Research Fund Program for Young Scholars. W.-L. Y. acknowledges NSFC under Grant Nos. 11474211 and 61674110.
\end{acknowledgments}

\appendix
\section{Proof of Proposition 2}\label{AppA}
We first assume $l_u$ is even, so that $l_u\geq 2$.
\par 1) $v=$ even.
\par For even $v$ we have $l_u+v-1=$ odd, so that $l_u+v-1\leq L-1$, and hence $h_1\neq0$ and $h_L\neq0$. Since $h_{l_u}=\cdots=h_{l_u+v-1}=0$, Eq.~(\ref{nbulk0_1}) leads to the \emph{possibility} (not necessarily) that
\begin{eqnarray}
B_{j+1}\neq 0,~~C_j\neq0,
\end{eqnarray}
where $j=\frac{l_u}{2},\frac{l_u}{2}+1, \cdots,\frac{l_u}{2}+\frac{v}{2}-1$, while all the rest of $\{B_i\}$ and $\{C_i\}$ must be zero. Inserting the above possibly nonzero variables into Eq.~(\ref{nbulk0_3})-(\ref{nboud0_2}), and using the fact that $h_{l}=0,~l\in T$, we obtain the following two sets of coupled equations, each of which consists of $\frac{v}{2}+1$ equations:
\begin{eqnarray}\label{EOMeeB}
-J^y_{\frac{l_u}{2}}B_{\frac{l_u}{2}+1}&=&0,\nonumber\\
-J^x_{\frac{l_u}{2}+j}B_{\frac{l_u}{2}+j}-J^y_{\frac{l_u}{2}+j}B_{\frac{l_u}{2}+1+j}&=&0,~~j=1,\cdots,\frac{v}{2}-1\nonumber\\
-J^x_{\frac{l_u}{2}+\frac{v}{2}}B_{\frac{l_u}{2}+\frac{v}{2}}-h_{l_u+v}D_{ \frac{l_u}{2}+\frac{v}{2}}&=&0,
\end{eqnarray}
and
\begin{eqnarray}\label{EOMeeC}
h_{l_u-1}A_{\frac{l_u}{2}}+J^x_{\frac{l_u}{2}}C_{\frac{l_u}{2}}&=&0,\nonumber\\
J^y_{\frac{l_u}{2}-1+j}C_{\frac{l_u}{2}-1+j}+J^x_{\frac{l_u}{2}+j}C_{\frac{l_u}{2}+j}&=&0,~j=1,\cdots,\frac{v}{2}-1\nonumber\\
J^y_{\frac{l_u}{2}+\frac{v}{2}-1}C_{\frac{l_u}{2}+\frac{v}{2}-1}&=&0.
\end{eqnarray}
Note that the above two sets of equations include the limiting cases with $l_u=2$ or $l_u+v=L$. Note also that the single nonvanishing field $h_{l_u+v}$ ($h_{l_u-1}$) appears in the last (the first) equation of set (\ref{EOMeeB}) [set (\ref{EOMeeC})].
\par Since we have assumed $J^x_j\neq 0$ and $J^y_j\neq 0$, $\forall j$, the above two sets of equations lead to
\begin{eqnarray}
B_{\frac{l_u}{2}+1}&=&B_{\frac{l_u}{2}+2}=\cdots=B_{\frac{l_u}{2}+\frac{v}{2}}=D_{\frac{l_u}{2}+\frac{v}{2}}=0,\nonumber\\
C_{\frac{l_u}{2}+\frac{v}{2}-1}&=&C_{\frac{l_u}{2}+\frac{v}{2}-2}=\cdots=C_{\frac{l_u}{2}}=A_{ \frac{l_u}{2}}=0.\nonumber
\end{eqnarray}
Thus, all the possibly nonvanishing $B$'s and $C$'s are actually zero for even $v$, i.e.,
\begin{eqnarray}
B_j=C_j=0,~\forall j.
\end{eqnarray}
Applying the above equation to Eqs.~(\ref{nbulk0_3})-(\ref{nboud0_2}) gives $v$ simple solutions for the eigenvector $V$:
\begin{eqnarray}\label{evenvV}
(0,\cdots,0,1^{(4j)},0,\cdots,0)^T,\nonumber\\
(0,\cdots,0,1^{(4j+1)},0,\cdots,0)^T,
\end{eqnarray}
where $j=\frac{l_u}{2},\frac{l_u}{2}+1,\cdots,\frac{l_u}{2}+\frac{v}{2}-1$ and the superscript $(i)$ in $1^{(i)}$ indicates that the $i$th element of $V$ is 1. According to the discussions at the end of Sec.~\ref{relation}, these $v$ simple solutions are actually the wavefunctions of the $v$ isolated EMZMs $d_{4j}$ and $d_{4j+1}$, $j=\frac{l_u}{2},\cdots,\frac{l_u}{2}+\frac{v}{2}-1$.
\par 2) $v=$ odd.
\par 2i) $v=1$.
\par In this case only $h_{l_u}$ is zero, so that $B_j=0,~\forall j$ and $C_{\frac{l_u}{2}}$ is the only possibly nonvanishing variable among the $C$'s. Based on these facts, we get from Eqs.~(\ref{nbulk0_3})-(\ref{nboud0_2}) the simple solution $(0,\cdots,1^{(2l_u)},\cdots)^T$ and the following two equations
\begin{eqnarray}
J^y_{\frac{l_u}{2}}C_{\frac{l_u}{2}}+h_{l_u+1}A_{\frac{l_u}{2}+1}&=&0,\nonumber\\
h_{l_u-1}A_{\frac{l_u}{2}}+J^x_{\frac{l_u}{2}}C_{\frac{l_u}{2}}&=&0,
\end{eqnarray}
which result in a nontrivial (unnormalized) solution
\begin{eqnarray}\label{Vv=1}
V=(\cdots,\mathcal{J}^{(x,h)}_{\frac{l_u}{2}},0,1^{(2l_u-1)},0,\mathcal{J}^{(y,h)}_{\frac{l_u}{2}},\cdots)^T,
\end{eqnarray}
where $\mathcal{J}^{(x,h)}_{i}$ and $\mathcal{J}^{(y,h)}_{i}$ are given by Eq.~(\ref{Jhua}).
\par 2ii) $v\geq 3$.
\par For odd $v\geq 3$, we have $l_u+v-1=$ even, so that $l_u+v-1\leq L$. The possibly nonvanishing variables are
\begin{eqnarray}
B_j&\neq&0,~j=\frac{l_u}{2}+1,\frac{l_u}{2}+2,\cdots,\frac{l_u}{2}+\frac{v-1}{2},\nonumber\\
C_j&\neq&0,~j=\frac{l_u}{2},\frac{l_u}{2}+1, \cdots,\frac{l_u}{2}+\frac{v-1}{2}.\nonumber
\end{eqnarray}
The corresponding two sets of equations read
\begin{eqnarray}\label{EOMeoB}
-J^y_{\frac{l_u}{2}}B_{\frac{l_u}{2}+1}&=&0,\nonumber\\
-J^x_{\frac{l_u}{2}+j}B_{\frac{l_u}{2}+j}-J^y_{\frac{l_u}{2}+j}B_{\frac{l_u}{2}+1+j}&=&0,~j=1,\cdots,\frac{v-3}{2}\nonumber\\
-J^x_{\frac{l_u}{2}+\frac{v-1}{2}}B_{\frac{l_u}{2}+\frac{v-1}{2}}&=&0,
\end{eqnarray}
and
\begin{eqnarray}\label{EOMeoC}
h_{l_u-1}A_{\frac{l_u}{2}}+J^x_{\frac{l_u}{2}}C_{\frac{l_u}{2}}&=&0,\nonumber\\
J^y_{\frac{l_u}{2}-1+j}C_{\frac{l_u}{2}-1+j}+J^x_{\frac{l_u}{2}+j}C_{\frac{l_u}{2}+j}&=&0,~j=1,\cdots,\frac{v-1}{2}\nonumber\\
J^y_{\frac{l_u}{2}+\frac{v-1}{2}}C_{\frac{l_u}{2}+\frac{v-1}{2}}+h_{l_u+v}A_{\frac{l_u}{2}+\frac{v+1}{2}}&=&0.
\end{eqnarray}
From the first set of equations we have $B_j=0,~\forall j$, which leads to $\frac{v+1}{2}$ simple EMZMs $d_{4j}$, $j=\frac{l_u}{2},\frac{l_u}{2}+1,\cdots,\frac{l_u}{2}+\frac{v-1}{2}$. However, in contrast to the case of even $v$, here we see that two nonvanishing fields $h_{l_u-1}$ and $h_{l_u+v}$ appear respectively in the first and last equation of the equation set (\ref{EOMeoC}). This results in two types of solutions: The trivial solution with $A_{\frac{l_u}{2}}= 0$ leads to $\frac{v-1}{2}$ isolated EMZMs $d_{4j+1}$, $j=\frac{l_u}{2},\frac{l_u}{2}+1,\cdots,\frac{l_u+v-3}{2}$. However, if $A_{\frac{l_u}{2}}\neq 0$, then we get an unnormalized \emph{nontrivial} solution
\begin{widetext}
\begin{eqnarray}
\label{Vvneq1}
V&=&\left[\cdots,\mathcal{J}^{(x,h)}_{\frac{l_u}{2}},0,1^{(2l_u-1)},0,0,0,\mathcal{J}^{(y,x)}_{\frac{l_u}{2}}, 0,0,0,\mathcal{J}^{(y,x)}_{\frac{l_u}{2}}\mathcal{J}^{(y,x)}_{\frac{l_u}{2}+1},\cdots,\prod^{\frac{l_u+v-3}{2}}_{i=\frac{l_u}{2}}\mathcal{J}^{(y,x)}_{i},0,\prod^{\frac{l_u+v-3}{2}}_{i=\frac{l_u}{2}}\mathcal{J}^{(y,x)}_{i}\mathcal{J}^{(y,h)}_{\frac{l_u+v-1}{2}},0,\cdots\right]^T,\nonumber\\
\end{eqnarray}
\end{widetext}
where $\mathcal{J}^{(y,x)}_i\equiv- J^y_i/J^x_{i+1} $. Note that $V$ is a real vector, so $V^T$ is the wavefunction of one of the EMZMs. We thus proved Proposition 2 for even $l_u$.
\par If the single consecutive sequence $T$ start with an odd $l_u$, then $l_u\geq 1$.
\par 1) $v=$ even, so that $l_u+v-1=$ even.
\par From Eq.~(\ref{nbulk0_1}), we have the possibility that
\begin{eqnarray}
B_j&\neq& 0,~C_j \neq  0,
\end{eqnarray}
where $j=\frac{l_u+1}{2},\frac{l_u+3}{2},\cdots,\frac{l_u+v-1}{2}$.
Inserting these variables into Eqs.~(5) and (6) and using the fact that $h_{l_u}=h_{l_u+1}=\cdots=h_{l_u+v-1}=0$, we get
\begin{eqnarray}\label{EOMoeB}
h_{l_u-1}D_{\frac{l_u-1}{2}}+J^y_{\frac{l_u-1}{2}}B_{\frac{l_u+1}{2}}&=&0,\nonumber\\
J^x_{\frac{l_u-1}{2}+j}B_{\frac{l_u-1}{2}+j}+J^y_{\frac{l_u-1}{2}+j}B_{\frac{l_u+1}{2}+j}&=&0,~j=1,\cdots,\frac{v}{2}-1\nonumber\\
J^x_{\frac{l_u+v-1}{2}}B_{\frac{l_u+v-1}{2}}&=&0,
\end{eqnarray}
and
\begin{eqnarray}\label{EOMoeC}
J^x_{\frac{l_u+1}{2}}C_{\frac{l_u+1}{2}}&=&0,\nonumber\\
J^y_{\frac{l_u-1}{2}+j}C_{\frac{l_u-1}{2}+j}+J^x_{\frac{l_u+1}{2}+j}C_{\frac{l_u+1}{2}+j}&=&0,~j=1,\cdots,\frac{v}{2}-1\nonumber\\
J^y_{\frac{l_u+v-1}{2}}C_{\frac{l_u+v-1}{2}}+h_{l_u+v}A_{\frac{l_u+v+1}{2}}&=&0.
\end{eqnarray}
Note that the above two sets of equations include the limiting cases with $l_u=1$ or $l_u+v-1=L$, for which the first equation in the first set or the last equation in the second set should be removed.
As in the case of even $l_u$, we have $B_j=C_j=0,~\forall j$, yielding $v$ simple EMZMs $ d_{2l_u-1},d_{2l_u+3},\cdots, d_{2l_u+2v-5} $ and $ d_{2l_u+2},d_{2l_u+6},\cdots, d_{2l_u+2v-2} $. For the special case of $l_u=1$ and $l_u+v-1=L$, we have $v=L$, then these $L$ simple EMZMs are just $d_1,d_4,d_5,d_8,\cdots,d_{L-3},d_{L}$, i.e., the MFs in the first row of the comb.
\par 2) $v=$ odd.
\par 2i) $v=1$.
\par In this case all the fields on the even sites are zero, so that $C_j=0,~\forall j$. This gives one simple solution $A_{\frac{l_u+1}{2}}=0$, resulting in a simple EMZM $d_{2l_u-1}$. In addition, $B_{\frac{l_u+1}{2}}$ is the only possibly nonvanishing variable among the $B$'s, which leads to the following two equations
\begin{eqnarray}
-J^x_{\frac{l_u+1}{2}}B_{\frac{l_u+1}{2}}-h_{l_u+1}D_{\frac{l_u+1}{2}}&=&0,\nonumber\\
-h_{l_u-1}D_{\frac{l_u-1}{2}}-J^y_{\frac{l_u-1}{2}}B_{\frac{l_u+1}{2}}&=&0.
\end{eqnarray}
The nontrivial solution is
\begin{eqnarray}
V=\left(\cdots,-\frac{J^y_{\frac{l_u-1}{2}}}{h_{l_u-1}},0,1^{(2l_u)},0,-\frac{J^x_{\frac{l_u+1}{2}}}{h_{l_u+1}},\cdots\right)^T.
\end{eqnarray}
\par 2ii) $v\geq3$.
\par For $v\geq 3$, we have $l_u+v-1\leq L-1$. The possibly nonvanishing variables are
\begin{eqnarray}
\label{Bj0_1}
B_j&\neq& 0,~j=\frac{l_u+1}{2},\cdots,\frac{l_u+v}{2},\\
\label{Cj0_1}
C_j&\neq& 0,~j=\frac{l_u+1}{2}, \cdots,\frac{l_u+v}{2}-1,
\end{eqnarray}
and hence
\begin{eqnarray}\label{EOMooB}
-h_{l_u-1}D_{\frac{l_u-1}{2}}-J^y_{\frac{l_u-1}{2}}B_{\frac{l_u+1}{2}}&=&0,\nonumber\\
-J^x_{\frac{l_u-1}{2}+j}B_{\frac{l_u-1}{2}+j}-J^y_{\frac{l_u-1}{2}+j}B_{\frac{l_u+1}{2}+j}&=&0,~j=1,\cdots,\frac{v-1}{2}\nonumber\\
-J^x_{\frac{l_u+v}{2}}B_{\frac{l_u+v}{2}}-h_{l_u+v}D_{\frac{l_u+v}{2}}&=&0,
\end{eqnarray}
and
\begin{eqnarray}\label{EOMooC}
J^x_{\frac{l_u+1}{2}}C_{\frac{l_u+1}{2}}&=&0,\nonumber\\
J^y_{\frac{l_u-1}{2}+j}C_{\frac{l_u-1}{2}+j}+J^x_{\frac{l_u+1}{2}+j}C_{\frac{l_u+1}{2}+j}&=&0,~j=1,\cdots,\frac{v-3}{2}\nonumber\\
J^y_{\frac{l_u+v-2}{2}}C_{\frac{l_u+v-2}{2}}&=&0.
\end{eqnarray}
The second set of equations give $C_j=0,~\forall j$, which results in $\frac{v+1}{2}$ simple EMZMs $d_{2l_{u}-1},d_{2l_{u}+3},\cdots,d_{2l_u+2v-3}$. The first set of equations have two types of solutions: 1) If $D_{\frac{l_u-1}{2}}=0$, then $B_j=0,~\forall j$, and hence we get $\frac{v-1}{2}$ simple EMZMs $d_{2l_u+2},\cdots,d_{2l_u+2v-4}$; 2) If $D_{\frac{l_u-1}{2}}\neq 0$, then we have obtain a nontrivial solution whose explicit expression will not be write down here for simplicity.
\section{Equation set for two coupled sequences}\label{AppC}
\par When two (odd) consecutive sequences $T_i$ and $T_{i+1}$ are coupled, the eigen-equations associated with them also get entangled. For even $l_{u_i}$, these coupled equations form an equation set
\begin{widetext}
\begin{eqnarray}\label{succ}
h_{l_{u_i}-1}A_{\frac{l_{u_i}}{2}}+J^x_{\frac{l_{u_i}}{2}}C_{\frac{l_{u_i}}{2}}&=&0,\nonumber\\
J^y_{\frac{l_{u_i}}{2}-1+j}C_{\frac{l_{u_i}}{2}-1+j}+J^x_{\frac{l_{u_i}}{2}+j}C_{\frac{l_{u_i}}{2}+j}&=&0,~j=1,\cdots,\frac{v_i-1}{2}\nonumber\\
J^y_{\frac{l_{u_i}+v_i-1}{2}}C_{\frac{l_{u_i}+v_i-1}{2}}+h_{l_{u_i}+v_i}A_{\frac{l_{u_i}+v_i+1}{2}}+J^x_{\frac{l_{u_{i+1}}}{2}}C_{\frac{l_{u_{i+1}}}{2}}&=&0,\nonumber\\
J^y_{\frac{l_{u_{i+1}}}{2}-1+j}C_{\frac{l_{u_{i+1}}}{2}-1+j}+J^x_{\frac{l_{u_{i+1}}}{2}+j}C_{\frac{l_{u_{i+1}}}{2}+j}&=&0,~j=1,\cdots,\frac{v_{i+1}-1}{2}\nonumber\\
J^y_{\frac{l_{u_{i+1}}+v_{i+1}-1}{2}}C_{\frac{l_{u_{i+1}}+v_{i+1}-1}{2}}+h_{l_{u_{i+1}}+v_{i+1}}A_{\frac{l_{u_{i+1}}+v_{i+1}+1}{2}}&=&0.
\end{eqnarray}
We can eliminate the $C$'s from the above equations to get a single equation for the three $A$'s:
\begin{eqnarray}\label{AAA}
\frac{\mathcal{J}^{(y,h)}_{\frac{l_{u_i}+v_i-1}{2}}}{\mathcal{J}^{(x,h)}_{\frac{l_{u_i}}{2}}} \prod^{\frac{l_{u_i}+v_i-3}{2}}_{j=\frac{l_{u_i}}{2}}\mathcal{J}^{(y,x)}_{j} A_{\frac{l_{u_i}}{2}}  - A_{\frac{l_{u_i}+v_i+1}{2}}+ \frac{\mathcal{J}^{(x,h)}_{\frac{l_{u_{i+1}}}{2}}}{ \mathcal{J}^{(y,h)}_{\frac{l_{u_{i+1}}+v_{i+1}-1}{2}} } \prod^{\frac{l_{u_{i+1}}+v_{i+1}-3}{2}}_{j=\frac{l_{u_{i+1}}}{2}}\frac{1}{\mathcal{J}^{(y,x)}_{j}} A_{\frac{l_{u_{i+1}}+v_{i+1}+1}{2}}=0.
\end{eqnarray}
\end{widetext}
\section{On the solution of Eq.~(\ref{detkfinal})}\label{AppB}
\par To find out the solutions to the quantization condition given by Eq.~(\ref{detkfinal}), we first investigate the properties of the function
\begin{eqnarray}
f(k,L)\equiv\frac{\sin 2k(L+2)}{\sin 2kL}.
\end{eqnarray}
As a primary illustration, we plot in Fig.~\ref{Fig4} $f(k,L=8)$ as a function of $k$. For \emph{real} $k$ and even $L$, the function $f(k,L)$ has the following several properties:
\begin{figure}
\includegraphics[width=.50\textwidth]{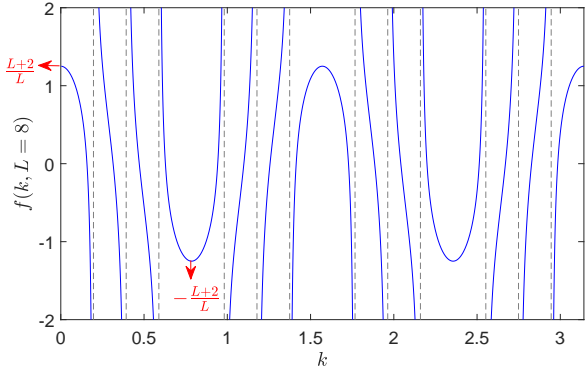}
\caption{The function $f(k,L)=\frac{\sin 2k(L+2)}{\sin 2kL}$ for $L=8$.}
\label{Fig4}
\end{figure}
\par 1). $f(k,L)$ has a period of $\frac{\pi}{2}$ in $k$,
\par 2). $f(k,L)$ is symmetric with respect to $k=\frac{m\pi}{4}$,~$m\in Z$
\par 3). $f(\frac{\pi}{4}-k,L)=-f(k,L)$,
\par 4). $\lim_{k\to\frac{m\pi}{2}}f(k)=\frac{L+2}{L},~\lim_{k\to\frac{(2m+1)\pi}{4} }f(k)=-\frac{L+2}{L}$,~$m\in Z$
\par Therefore, we can focus on the solutions of Eq.~(\ref{detkfinal}) on the interval $k\in[0,\frac{\pi}{4}]$, where $\frac{2L}{4}=\frac{L}{2}$ (recall that the dispersion $\lambda_k$ has four branches) real solutions is expected to be found for certain parameters. For simplicity, we avoid discussing the limiting case with $\gamma=\frac{L+2}{L}$.
\par In general, the solution $k=k_0+ik_1$ could be complex. As an example, consider the parameter range with $|\gamma|>\frac{L+2}{L}$, then from Fig.~\ref{Fig4} we see that there are only $3=\frac{L}{2}-1$ \emph{real} roots of $k$ on $[0,\frac{\pi}{4}]$. So we must have one additional \emph{complex} root of $k$. To determine the most general form of a complex root, we first note that $\lambda_k$ must be purely imaginary, so that
\begin{eqnarray}
\delta_k=J^2_x[1+\gamma^2+ \gamma(e^{i4k_0}e^{-4k_1}+e^{-i4k_0}e^{4k_1})]
\end{eqnarray}
must be real, which gives
\begin{eqnarray}
 \sin 4k_0(e^{-4k_1}-e^{4k_1})=0.
\end{eqnarray}
For real $k$ with $k_1=0$, this equation is satisfied automatically; for complex $k$ with $k_1\neq 0$, we must have
\begin{eqnarray}
 \sin 4k_0=0\to 4k_0=m\pi,~m\in Z
\end{eqnarray}
then the complex $k$ can be written
\begin{eqnarray}\label{kpi4}
k=\frac{m}{4}\pi+ik_1.
\end{eqnarray}
\par By inserting Eq.~(\ref{kpi4}) into Eq.~(\ref{detkfinal}), we obtain the condition equation for $k_1$
\begin{eqnarray} \label{sinh}
 \gamma&=& (-1)^{m-1}g(k_1,L),
\end{eqnarray}
where $g(k,L)=\frac{\sinh 2k(L+2)}{\sinh 2kL}$. It is easy to see that:
\par 1) $g(k,L)=g(-k,L)$,
\par 2) $\lim_{k\to0}g(k,L)=\frac{L+2}{L}$,
\par 3) $\frac{\partial g(k,L)}{\partial k}\geq0$ for $k\geq 0$, so that $g(k,L)>\frac{L+2}{L}$ for $k>0$.
\par Since we have assumed $\gamma>0$, Eq.~(\ref{sinh}) has real solutions only for $m=$ odd, say $m=1$, so that the complex root can be written
\begin{eqnarray}
k=\frac{\pi}{4}+ik_1,
\end{eqnarray}
and the condition given by Eq.~(\ref{sinh}) becomes
\begin{eqnarray}\label{gammagL}
\gamma&=&g(k_1,L)> \frac{L+2}{L}
\end{eqnarray}
for $k_1 > 0$. The inequality holds since $g(k,L)$.
\par Thus, the solutions of Eq.~(\ref{detkfinal}) can be summarized as follows:
\par 1). For $\gamma<\frac{L+2}{L}$, it is impossible to have any complex solution on $[0,\frac{\pi}{4}]$, and hence all the $\frac{L}{2}$ solutions are real. We can also see this from Fig.~\ref{Fig4}.
\par 2). For $\gamma>\frac{L+2}{L}$, there are $\frac{L}{2}-1$ real solutions on $[0,\frac{\pi}{4}]$. In addition, there exists a single complex solution~\cite{Lieb,finiteIsing}
\begin{eqnarray}
\tilde{k}=\frac{\pi}{4}+ i\tilde{k}_1,
\end{eqnarray}
with $\tilde{k}_1$ being the solution of Eq.~(\ref{gammagL}) on $[0,\infty)$. We can choose $\tilde{k}_1>0$ since $g(k_1,L)$ is an even function of $k_1$. The four eigenvalues $\lambda^{(\pm)}_{\tilde{k},\pm}$ corresponding to the complex solution $\tilde{k}$ are still given by Eq.~(\ref{dispersion}), but with $\delta_k$ replaced by
\begin{eqnarray}
\tilde{\delta}_{\tilde{k}}=J^2_x(1+\gamma^2-2\gamma\cosh 4\tilde{k}_1).
\end{eqnarray}
\par For real $k$, we always have $\delta_k=J^2_x[(\gamma+\cos 4k)^2+(\sin 4k)^2]\geq 0$, where the equality holds for $\gamma=1$ and $k=\frac{\pi}{4}$. But $f(k,L)$ approaches $-\frac{L+1}{L}$ as $k\to \frac{\pi}{4}$, indicating that $k=\frac{\pi}{4}$ is not a solution to Eq.~(\ref{detkfinal}). We thus have $\delta_k>0$ for all real solutions of Eq.~(\ref{detkfinal}). For the complex solution $\tilde{k}=\frac{\pi}{4}+i\tilde{k}_1$, we can write $\tilde{\delta}_{\tilde{k}}=J^2_x[(\gamma \cosh 4\tilde{k}_1-1)^2-(\gamma\sinh4\tilde{k}_1)^2]$, with $\gamma>\frac{L+2}{L}$. Note that for $\tilde{k}_1>0$ we have $\cosh 4\tilde{k}_1>1$, and $\sinh 4\tilde{k}_1>0$. From
\begin{eqnarray}\label{cosh-sinh}
&&(\gamma \cosh 4\tilde{k}_1-1)-\gamma\sinh4\tilde{k}_1=g(\tilde{k}_1,L) e^{-4\tilde{k}_1}-1\nonumber\\
&=&2\sinh4\tilde{k}_1\frac{e^{-4\tilde{k}_1}}{e^{4\tilde{k}_1 L}-1}>0,\nonumber
\end{eqnarray}
we see that $\tilde{\delta}_{\tilde{k}}>0$.


\begin{thebibliography}{99}
\bibitem{OEdeg} Accidental degeneracies may occur in certain cases. For example, if $H^{\mathrm{(MF)}}_{\mathrm{PBC},\sigma}$ possesses one or several nonzero eigenenergies that are degenerate. Another example is the case in which the sub-ground states in the two subspaces are degenerate (e.g., in the quantum Ising chain with vanishing transverse field under the PBC).
\end{thebibliography}
\end{document}